\documentclass[conference]{IEEEtran}
\IEEEoverridecommandlockouts  

\usepackage[T1]{fontenc} 
\usepackage[normalem]{ulem} 

\usepackage[usenames,dvipsnames,svgnames,x11names]{xcolor}

\usepackage{moreverb}

\usepackage[nounderscore]{syntax}

\usepackage[cmex10]{amsmath}
\usepackage{amssymb}
\usepackage{mathtools}
\usepackage{amsthm}
\usepackage{amsfonts}

\usepackage{gensymb}

\usepackage{subfig} 
\usepackage[export]{adjustbox}

\usepackage{algorithmicx}
\usepackage{algpseudocode}
\usepackage[ruled]{algorithm}
\definecolor{light-gray}{gray}{0.75}
\definecolor{dark-gray}{gray}{0.5}
\algrenewcommand{\algorithmiccomment}[1]{\hskip3em{{\footnotesize \textcolor{light-gray}{$\blacktriangleright$}~\textcolor{dark-gray}{#1}}}}
\usepackage{multirow} 
\usepackage{tabularx}
\usepackage{rotating} 
\usepackage{booktabs} 
\usepackage{colortbl} 
\usepackage{tablefootnote} 

\usepackage[pdftex,colorlinks=true,urlcolor=blue,citecolor=red]{hyperref}

\usepackage{xspace}

\usepackage{enumitem}

\usepackage{balance}

\begin{document}

\title{Bounding Queue Waits for Fair Scheduling of Bursty HPC Workloads via Hierarchical Algorithms}

\author{
\IEEEauthorblockN{Karaen Senthilkumar$^{1}$, Suman Raj$^{1,2}$, Maxime Gonthier$^{3}$, Kyle Chard$^{1,2}$, Donpaul Stephens$^{4}$, Ian Foster$^{2,1}$}
\IEEEauthorblockA{$^{1}$Department of Computer Science, University of Chicago, IL, USA\\
$^{2}$Data Science and Learning Division, Argonne National Laboratory, Lemont, IL, USA\\
$^{3}$Inria Centre at the University of Bordeaux, Bordeaux, France\\
$^{4}$AirMettle, Inc., Houston, TX, USA\\
Email: \{karaen, sumanraj, chard\}@uchicago.edu, maxime.gonthier@inria.fr, donpaul@airmettle.com, foster@anl.gov}
}

\maketitle
\thispagestyle{plain}
\pagestyle{plain}

\begin{abstract}
HPC clusters are shared among many projects, each holding an allocation of node-hours, and ensuring fairness is becoming harder. Foundation models workflows and automated pipelines have made job submissions bursty and concentrated among a few heavy users, so occasional users absorb queueing delays created by frequent ones, with no indication of how long their jobs will wait. Existing schedulers either forget past usage, protect only the job at the head of the queue, or assume that running jobs can be preempted. We propose HyDRA, a Hybrid Deadline and Ranking-Aware scheduler that enforces fairness among users while bounding the queue wait of every job. HyDRA assigns each job a deadline on arrival based on its size, the priorities of its project and user, the current backlog, and how far its project and user are from their entitled share. It ranks waiting jobs by a priority score that grows with waiting time, serves jobs close to their deadlines first, and backfills a job only when no deadline, or fairness limit is violated. We evaluate HyDRA on workloads derived from a year of ALCF Polaris job traces, against conventional schedulers and variants inspired by PCS, Themis, and the PBS job-sort formula used at ALCF. HyDRA achieves $2\times$ the project-level Jain's fairness of the best baseline, starts 47\% of jobs by their deadlines compared with at most 36\% for the baselines even for largest workloads, and lowers the 95th-\%ile bounded slowdown by 4.2$\times$ relative to the next best baseline, at a moderate cost in utilization.
\end{abstract}

\section{Introduction}\label{sec:intro}

Most HPC facilities, from leadership computing centers such as the Argonne Leadership Computing Facility (ALCF) to university clusters, allocate compute time to projects, which is consumed by jobs submitted by individual project users. At the job level, priority policies and backfilling determine which jobs run next. When HPC workloads consisted of long, planned batch campaigns, this was a reasonable approach; however, the adoption of machine learning models in scientific computing has changed submission behavior. Hyperparameter sweeps, inference campaigns, and checkpoint-restart training runs often arrive in dense bursts separated by relatively quiet periods. Arrivals are now far more variable, and that variability is concentrated among a small number of users.

\begin{figure}[t]
\centering
\subfloat[Polaris: users ranked by core-hours consumed.]{%
  \includegraphics[width=0.5\columnwidth]{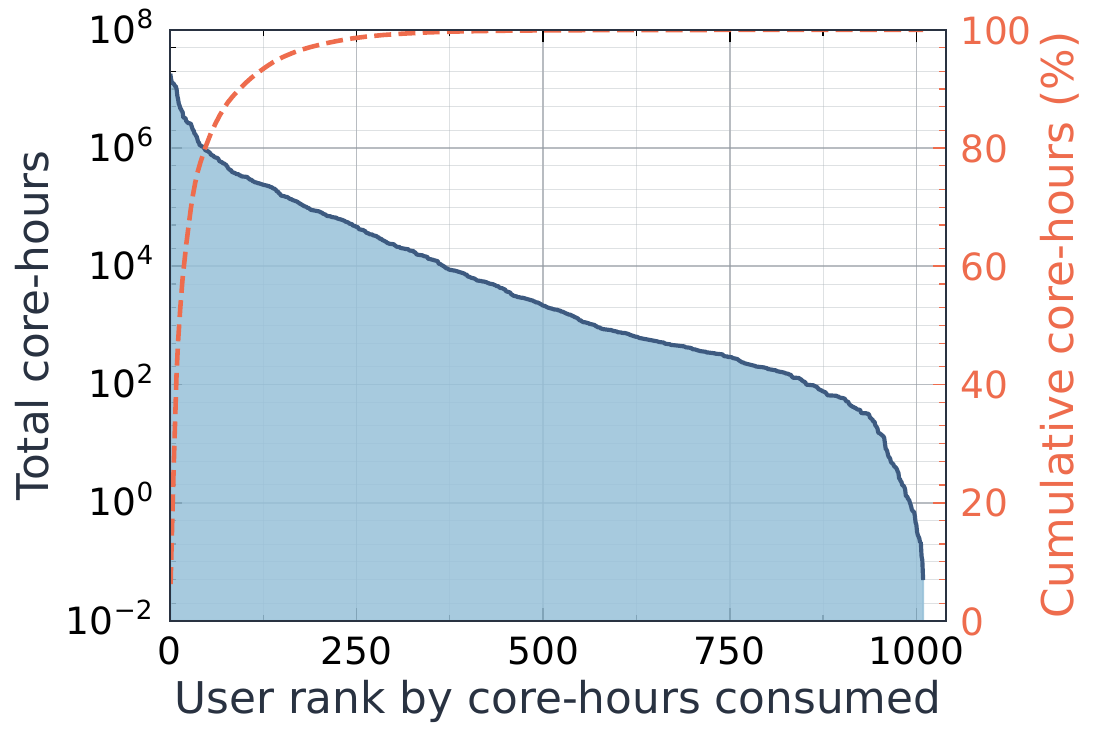}%
  \label{fig:motivation-pareto-polaris}}
\subfloat[Polaris: burstiness $B$ against number of jobs submitted.]{%
  \includegraphics[width=0.48\columnwidth]{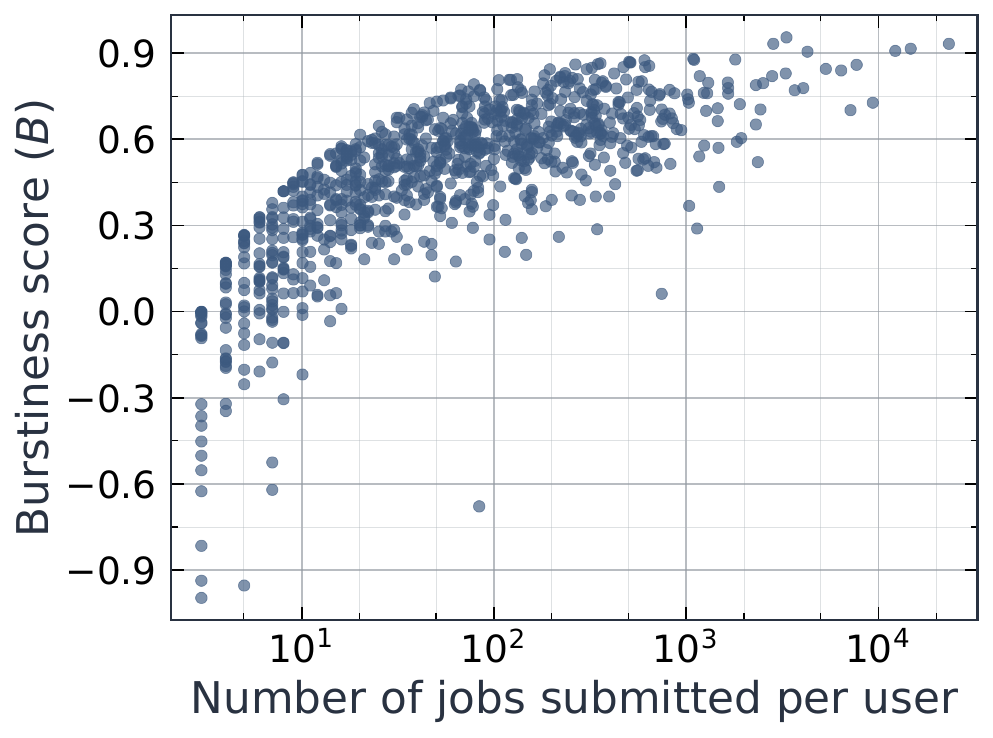}%
  \label{fig:motivation-burstiness-polaris}}\\
\subfloat[Midway3: users ranked by core-hours consumed.]{%
  \includegraphics[width=0.5\columnwidth]{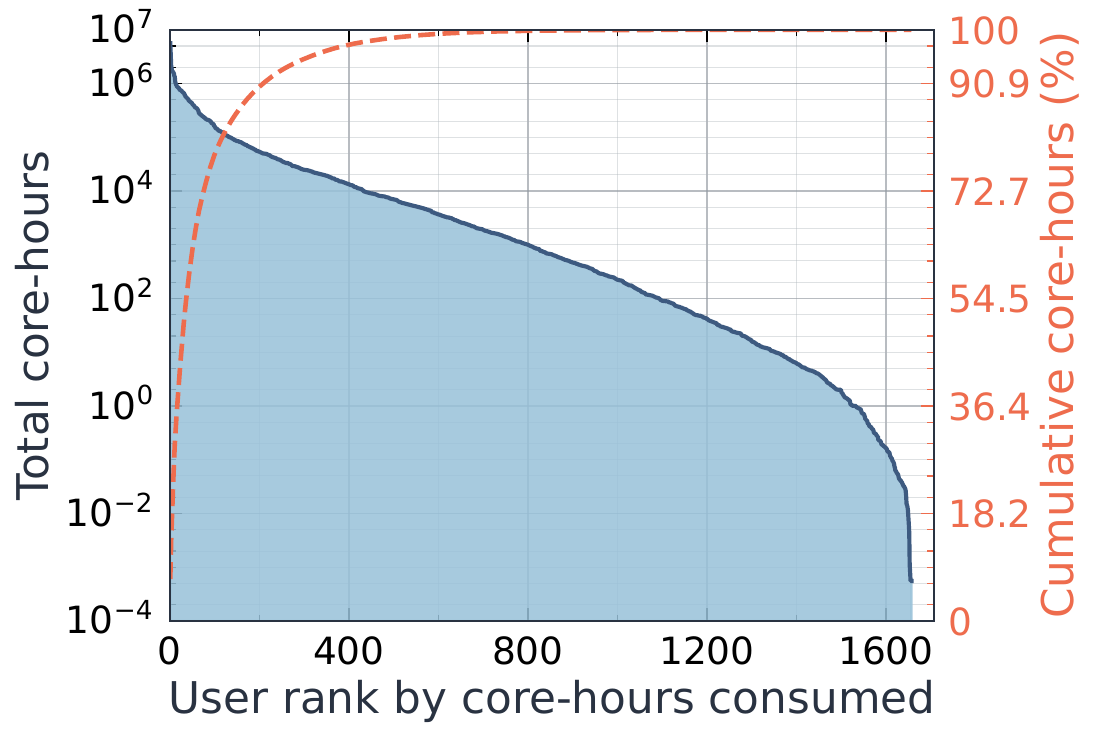}%
  \label{fig:motivation-pareto-midway}}
\subfloat[Midway3: burstiness $B$ against number of jobs submitted.]{%
  \includegraphics[width=0.48\columnwidth]{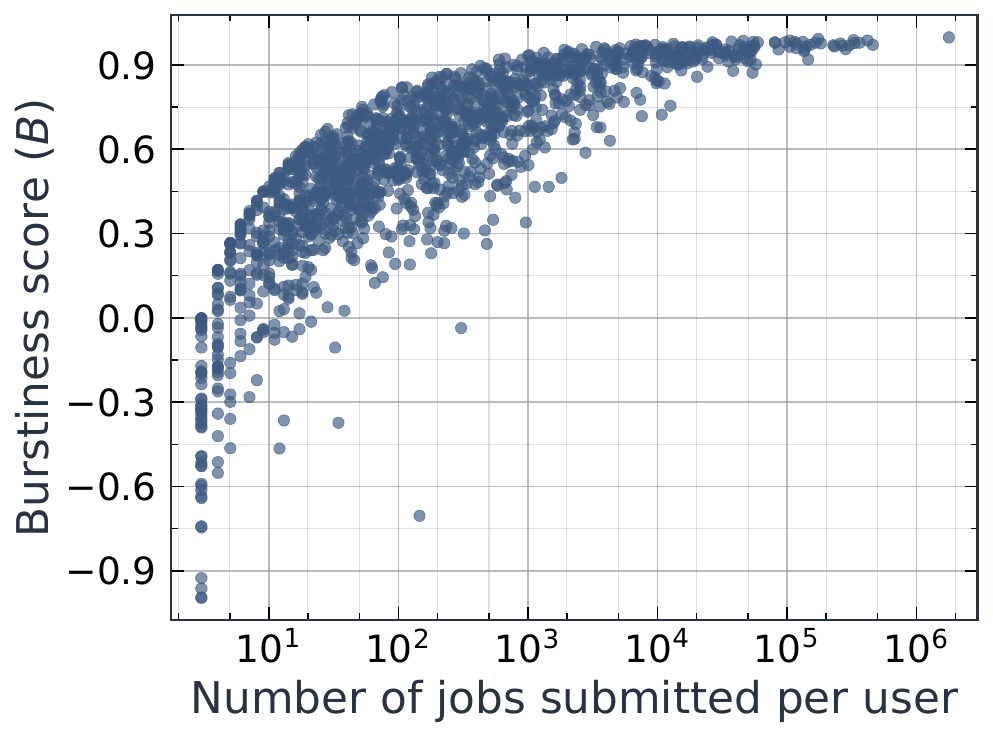}%
  \label{fig:motivation-burstiness-midway}}
\caption{On both Polaris and Midway3, a few users consume most of the resources (a, c), and the users who submit the most jobs are also the burstiest (b, d).}
\label{fig:motivation}
\end{figure}

\begin{figure}[t]
\centering
\includegraphics[width=\columnwidth]{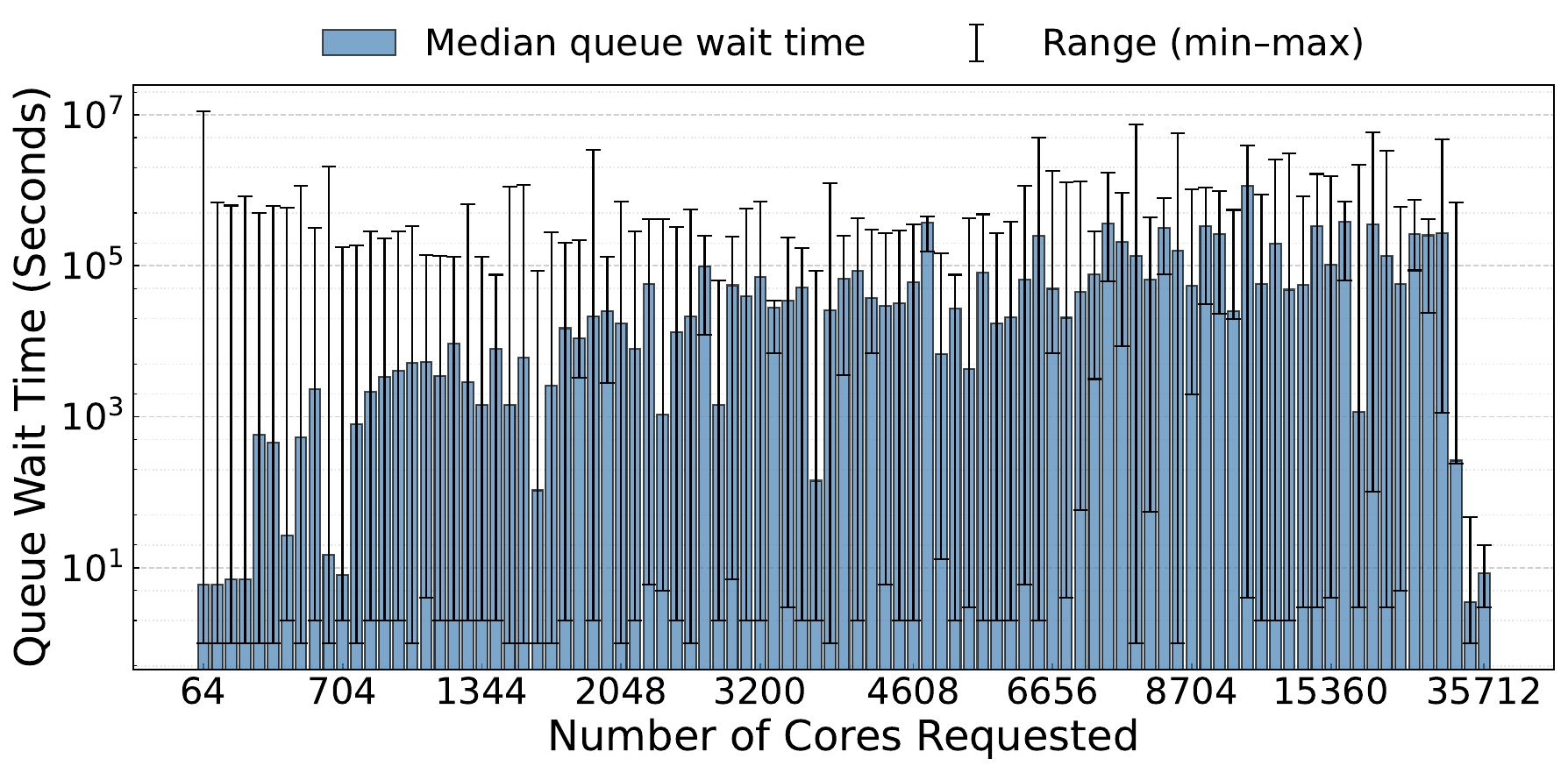}
\caption{Mean and range of queue times versus the number of CPU cores requested for Polaris.}
\label{fig:polaris-qtime-cores-requested}
\end{figure}

Figure~\ref{fig:motivation} shows how uneven this sharing has become. We analyze traces from a year of job records on ALCF Polaris: roughly $20\%$ of users consume more than $95\%$ of the core-hours (Fig.~\ref{fig:motivation-pareto-polaris}), and consumption on the University of Chicago's RCC Midway3 cluster is similarly concentrated (Fig.~\ref{fig:motivation-pareto-midway}). The heaviest users are also the burstiest. On both systems, the Goh--Barab\'asi burstiness parameter $B$~\cite{Goh_2008} rises with the number of jobs a user submits (Figs.~\ref{fig:motivation-burstiness-polaris} and~\ref{fig:motivation-burstiness-midway}). Moreover, because the queue time on Polaris also grows with the number of cores requested (Fig.~\ref{fig:polaris-qtime-cores-requested}), the delay created by a burst is not shared evenly. Hence, we observe that a heavy user's burst fills the queue, and a large job from an occasional user must wait behind the burst before it can be placed. Users who submit rarely and at scale end up absorbing the delay generated by users who submit often.

Preventing bursty users from undermining fairness for occasional users is challenging for three major reasons. First, HPC jobs are rigid and usually non-preemptible. A tightly coupled job holds all of its nodes until it finishes and generally cannot be paused or resized without application support~\cite{Fan2022Hybrid}.

Second, fairness is hierarchical. Projects have allocation budgets that must not be exceeded, and within each project, users have project-specific priorities, so the scheduler must prevent one project's burst from consuming another's share and one user's burst from crowding out teammates.
Third, waiting-time bounds are hard to guarantee.
Under bursty load, and with walltime requests that overestimate runtimes~\cite{Mualem2001Backfilling,Tsafrir2007Backfilling}, a bound that is too tight is routinely missed, while a bound that is too loose is of no use to users.
Moreover, enforcing a bound must not leave the machine idle.

Existing mechanisms for fairness do not fully address this problem.
Fair-share schedulers can use past usage to regulate fairness, but only with finite memory.
For instance, Slurm's multifactor plugin decays historical consumption with a configurable parameter~\cite{RCACFairshare2025}, so a user who bursts, goes quiet, and returns has their fairness index restored. EASY backfilling~\cite{Lifka1995EASY,Mualem2001Backfilling}, commonly layered over First-Come-First-Served (FCFS) ordering in HPC clusters, raises utilization by letting a short job run ahead of the queue head, provided it does not delay the head job's reservation. Thus, a job can be repeatedly overtaken because the mechanism does not account for how long it has already waited.

As Table~\ref{tab:rw-comparison} summarizes, no existing scheduler enforces project-level allocation budgets, user-level fairness within each project, and per-job waiting-time bounds for rigid, non-preemptible jobs.

We present HyDRA, a hybrid deadline and ranking-aware hierarchical scheduler that explicitly separates fairness across projects from fairness among users within each project, while enforcing waiting-time guarantees for non-preemptible jobs. At the project level, the scheduler accounts for each project's allocation budget. Within a project, service is allocated among users according to the desired fairness policy. Most importantly, each waiting job is assigned a deadline derived from this fairness objective. With our model, backfilling can now respect fairness as a lower-priority job may be backfilled only when doing so preserves the deadlines of waiting jobs, respects the project's remaining allocation budget, and keeps each user's service deficit within a specified tolerance. 

In this paper, we make the following key contributions: 

\begin{enumerate}[leftmargin=*]
    \item We first characterize job traces from ALCF Polaris (\S~\ref{sec:intro}). We use this to motivate and define the problem of scheduling rigid jobs submitted by the users of many projects to a shared HPC cluster while ensuring fairness (\S~\ref{sec:problem}).
    \item We propose the architecture of HyDRA, a hierarchical scheduling framework, that enforces precise project allocation budgets while optimizing fairness between users within a project.(\S~\ref{sec:architecture}).
    \item We propose fairness-aware hierarchical scheduling heuristics for HyDRA that derive the entitlements and priorities of projects, users, and jobs at three levels of hierarchy, rank waiting jobs by a dynamic priority score that grows with their waiting time, and assign each job an adaptive deadline from its size, its priority, and the current backlog (Section~\ref{sec:scheduling}). We also show mathematically that, with this score, any two waiting jobs exchange rank at most once (\S~\ref{sec:analysis}).
    \item We evaluate HyDRA against six baselines, including SOTA schedulers and the production policies of ALCF, on workloads sampled from ALCF Polaris trace. Under heavy load, HyDRA achieves twice the project-level Jain's index of the best baseline, starts 47\% of jobs by their deadline compared with at most 36\% for the baselines, and lowers the 95th-percentile bounded slowdown by 4.2$\times$, at a modest cost of utilization. Its backlog term keeps deadlines achievable for longer under bursts, which improves both its fairness and its on-time rate (\S~\ref{sec:evaluation}).
\end{enumerate}

\section{Related Work}\label{sec:related}
We review prior HPC work in three areas: (1) scheduling, fair-share, and fair queueing; (2) queue-wait prediction and wait-time bounds; and (3) deadlines, slowdown, and starvation.
Table~\ref{tab:rw-comparison} compares state-of-the-art (SOTA) approaches on the four properties that HyDRA combines: support for rigid HPC jobs, project-level entitlements, user-level fairness, and a per-job wait bound. None of the others provides all four.

\begin{table}[t]
\centering
\scriptsize
\caption{HyDRA compared to SOTA. \checkmark: supported; (\checkmark): soft or partial; --: not supported. The last column gives the source of a per-job wait bound, if any.}
\label{tab:rw-comparison}
\setlength{\tabcolsep}{1pt}
\begin{tabularx}{\columnwidth}{@{}>{\raggedright\arraybackslash}Xcccl@{}}
\toprule
 & \textbf{Rigid} & \textbf{Project} & \textbf{User} & \textbf{Per-job} \\
\textbf{Approach} & \textbf{HPC jobs} & \textbf{entitlement} & \textbf{fairness} & \textbf{wait bound} \\
\midrule
EASY backfilling~\cite{Lifka1995EASY,Mualem2001Backfilling} & \checkmark & -- & -- & Arrival order \\
Selective reservation~\cite{Srinivasan2002Selective} & \checkmark & -- & -- & Global threshold \\
Learned schedulers~\cite{Wang2025HRLSched} & \checkmark & -- & -- & -- \\
Fair-share (Slurm)~\cite{RCACFairshare2025} & \checkmark & (\checkmark)\textsuperscript{a} & (\checkmark)\textsuperscript{a} & -- \\
WF$^2$Q, H-PFQ~\cite{Bennett1996WF2Q,Bennett1996HPFQ} & -- & \checkmark & \checkmark & Divisible service only \\
UWFQ~\cite{Kazemaks2025UWFQ} & -- & -- & \checkmark & -- \\
Queue-wait predictors~\cite{Nurmi2008QBETS,Okafor2026QueueWait} & \checkmark & -- & -- & Predicted, not enforced \\
Deadline scheduling~\cite{Liu1973EDF,Gonthier2025Deadline} & (\checkmark)\textsuperscript{b} & -- & -- & Given as input \\
\midrule
 \rowcolor{green!25} \textbf{HyDRA (our work)} & \checkmark & \checkmark
 & \checkmark & Assigned at arrival \\
\bottomrule
\end{tabularx}
\vspace{2pt}
{\raggedright\scriptsize
\textsuperscript{a}Relative shares computed from decayed usage; no hard budget.
\textsuperscript{b}EDF's optimality assumes preemption; Gonthier et al.\ interleave critical and non-critical tasks instead of preempting them.
\par}
\end{table}

\subsection{HPC Scheduling, Fair-Share, and Fair Queueing}
Most HPC centers combine a multifactor priority queue with backfilling~\cite{
RCACFairshare2025,
Allcock2017Argonne}. Backfilling variants differ in which jobs they protect. EASY backfilling protects only the job at the head of the queue~\cite{Lifka1995EASY}, conservative backfilling protects every job in arrival order~\cite{Mualem2001Backfilling}, and selective reservation protects a job only once its slowdown passes a global threshold~\cite{Srinivasan2002Selective}.
SOTA schedulers learn which job to run next~\cite{Wang2025HRLSched}
%
or account for burst buffers, multiple resources, power caps, and heterogeneous or changing cluster capacity~\cite{
Yoshida2025VAHRM,Benoit2026Variable}, but they still optimize averages from a cluster perspective and promise nothing to any single job. 
In contrast, HyDRA protects every waiting job with its own deadline, and it backfills a job only if every deadline still holds, the project stays within its budget, and the user's service deficit stays within its tolerance.

In HPC systems, fairness between users typically comes from fair-share, which lowers a user's priority according to their recent usage~\cite{RCACFairshare2025}.
%
Its outcome depends heavily on configuration~\cite{Klusacek2025FairSharingSim}, and because past usage decays, a user who bursts and then goes quiet soon gets their priority back. Metrics such as fair start time~\cite{Sabin2004JobFairness}, Jain's index~\cite{Jain1984Fairness}, and packing-aware fairness measures~\cite{Goponenko2022Metrics} can show the damage afterwards, but they do not prevent it. HyDRA instead enforces project allocations as hard budgets.

Fair queueing offers what fair-share lacks~\cite{Bennett1996WF2Q}: each flow receives a proportional share with bounded delay. Some~\cite{Bennett1996HPFQ} extends this to a hierarchy form, much like our projects and users, while others~\cite{Kazemaks2025UWFQ} adapt it to bursty users. All of these assume that service can be divided into small pieces, such as packets in networking or short tasks, so that a user who falls behind can be repaid quickly. A rigid HPC job holds its nodes until it finishes, which leaves the scheduler a single knob of execution start time. HyDRA keeps project-to-user hierarchy but bounds execution start times rather than service lag.

\subsection{Queue-Wait Prediction and Wait-Time Bounds}
Users have long wanted to know how long their jobs will wait in the queue. Early work predicted queue times using various methods, including statistical bounds on observed waits~\cite{Nurmi2008QBETS}. More recently, supervised models have been trained on job data from ALCF~\cite{Okafor2026QueueWait}, and runtime predictors aim to make backfilling reservations more precise~\cite{Gaussier2015Backfilling}, although better predictions do not always produce better schedules~\cite{Kumar2026TraceDriven}. In all of these works, however, the prediction is never fed back into the scheduler.

HyDRA, in contrast, predicts a deadline for each incoming job and then schedules the job to meet it. Conservative backfilling also gives each job a start-time bound~\cite{Mualem2001Backfilling}, but it computes that bound from an arrival-ordered queue, so a job that arrives behind a burst inherits the burst's delay regardless of which user submitted it. HyDRA's deadline likewise accounts for the current cluster capacity and the job's position in the queue, but it also weighs the user's priority and the job's size, so the bound reflects the intended fairness between users rather than arrival order alone.

\subsection{Deadlines, Slowdown, and Starvation}

Classical deadline scheduling takes deadlines as provided by the user.
EDF is optimal on a single preemptive processor~\cite{Liu1973EDF}, and Gonthier et al.~\cite{Gonthier2025Deadline} avoid reserving or preempting resources by interleaving critical and non-critical tasks. In ML clusters, Chronus~\cite{Gao2021Chronus} and DeepPlus~\cite{Xi2026DeepPlus} likewise try to meet completion deadlines that users provide. In HyDRA, by contrast, jobs arrive without deadlines, and the scheduler assigns each one an estimated deadline. Per-job delay is usually measured by stretch and maximum flow in theory~\cite{Bender1998Stretch,Canon2022FlowTime} and by bounded slowdown in HPC~\cite{Feitelson2001Metrics}.
Favoring small jobs improves average stretch and slowdown but can be unfair to large jobs~\cite{DePellegrini2024CoFair}, and HyDRA makes the weight given to job size tunable. Some schedulers fight starvation with aging~\cite{Yoshida2025VAHRM}, but aging alone guarantees nothing, since how fast a job moves up depends on the other jobs in the queue. HyDRA therefore uses aging to order the queue and relies on deadlines as a backstop.

\section{HyDRA Framework}\label{sec:framework}

In this section, we formulate the problem of scheduling rigid jobs submitted by users of many projects to a shared HPC cluster while ensuring fairness and describe the architecture of HyDRA scheduler we propose to solve it.

\subsection{Problem Formulation}\label{sec:problem}

We assume an HPC facility, such as the ALCF, in which a cluster is shared by many projects, with users submitting jobs to a common batch scheduler. Our system model comprises three key components: the cluster that runs jobs, the projects to which resources are allocated, and the users who submit jobs under each project. The cluster has $N$ nodes, each characterized by its CPU cores, memory, and network bandwidth. We denote by $C(t)$ the node-time it can deliver during the time interval $[0,t]$. For this paper, we consider homogeneous CPU nodes. Each project $p_j\in P$ has an \textit{allocation budget} $B_j$, the node-hours it may consume over an allocation period, and a set of users $\mathcal{U}_j$ who submit jobs against it, where $u_{ij}$ denotes user $i$ of project $p_j$. A person who works on several projects is modeled as a separate user for each project. Users may submit jobs online, often in bursts. For example, a user running a hyperparameter sweep may submit hundreds of jobs within minutes, whereas another user on a different project submits a single large simulation. Let $S_{ij}(t)$ denote the cumulative service, in the same units as the budgets, that user $u_{ij}$ has received up to time $t$, so that project $p_j$ has received $S_j(t)=\sum_i S_{ij}(t)$ and the cluster has delivered $S(t)=\sum_j S_j(t)$. Two constraints hold at all times: a project cannot exceed its budget, $S_j(t)\le B_j$, and the cluster cannot deliver more than its capacity, $S(t)\le C(t)$. 

Each job submitted by a user is represented as a tuple $J_k=(a_k,\tau_k,n_k,p(J_k),u(J_k))$, where $a_k$ is its arrival time, $\tau_k$ its requested walltime, $n_k$ the number of nodes it requests, and $p(J_k)$ and $u(J_k)$ identify the project and the user that submitted it. Its \textit{service demand} $q_k=n_k\tau_k$ is the node-time that the job may consume. Jobs are rigid and non-preemptible: a job starts only when all of its requested nodes are free, and holds them until it completes. If job $J_k$ starts at time $s_k$ and runs for $r_k$, it completes at $e_k=s_k+r_k$ after waiting $\omega_k=s_k-a_k$ in the queue. The runtime $r_k$ is unknown when the job arrives and is typically well below $\tau_k$. Each submitted job is assigned a \textit{deadline} $D_k=a_k+\Omega_k$ when it arrives, i.e., the latest time by which the job should start, where $\Omega_k$ bounds its waiting time. Unlike a deadline supplied by the user, $\Omega_k$ is computed by the scheduler using Eqn.~\ref{eqn:deadline}. A job meets its deadline if $s_k\le D_k$, or equivalently $w_k\le\Omega_k$.

We consider three levels of hierarchy. The facility assigns each project $p_j$ a \textit{weight} $w_j$ that reflects its relative priority, and the project assigns each of its users $u_{ij}$ a weight $w_{ij}$ that reflects the user's priority within the project. These weights determine the share of the cluster that each project and user is \textit{entitled} to (\textit{Level-2}). If a project does not assign weights to its users, all of its users have the same entitlement (\textit{Level-1}). A job $J_k$ may also carry its own weight $w_k$, e.g., from the queue to which it is submitted (\textit{Level-3}). A job thus inherits the priority of its project at Level-1, the combined priority of its project and user at Level-2, and the combined priority of its project, its user, and itself at Level-3. We normalize these weights to a fixed range, which we discuss in detail in \S~\ref{sec:entitlement}.

\textbf{Problem.} \textit{For a stream of rigid jobs submitted online by the users of multiple projects, schedule them on the cluster such that: (i) no project exceeds its allocation budget, (ii) each job starts no later than its deadline $D_k$, (iii) the service delivered to each project and user tracks its entitlement, and (iv) the cluster utilization remains high.}

\begin{figure}[t]
\centering
\includegraphics[width=0.95\columnwidth]{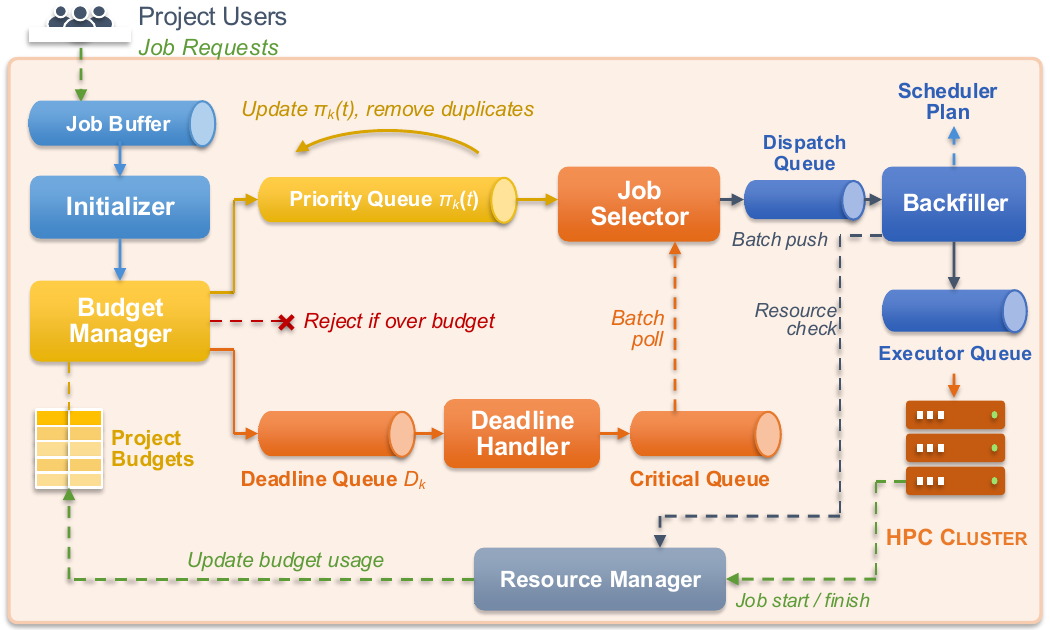}
\caption{HyDRA scheduler architecture.}
\label{fig:hydra-architecture}
\end{figure}

\subsection{Architecture}\label{sec:architecture}
Fig.~\ref{fig:hydra-architecture} shows the architecture of the HyDRA scheduler. Users of each project submit job requests to the scheduler, which places them in a \textit{job buffer} in FCFS order. An \textit{initializer} thread takes one job $J_k$ at a time from this buffer, computes its service demand $q_k$, its priority $\rho_k$, and its initial priority score $\pi_k(a_k)$, and assigns its deadline $D_k$. The job is then passed to the \textit{budget manager}, which looks up the remaining budget $B_j-S_j(t)$ of its project in the \textit{project budgets} table and rejects the job request if its service demand $q_k$ exceeds this amount. On a successful check, the budget manager inserts the job simultaneously into two queues: a \textit{priority queue} ordered by the dynamic priority score $\pi_k(t)$, and a \textit{deadline queue} ordered by the deadline $D_k$. Since $\pi_k(t)$ grows as a job waits, the scores in the priority queue are updated periodically. 

A \textit{deadline handler} monitors the head of the deadline queue and moves each job that becomes \textit{deadline-critical}, i.e., $D_k-t\le\delta$ for a slack $\delta\ge0$, into an FCFS \textit{critical queue}. The \textit{job selector} then pushes jobs into an FCFS \textit{dispatch queue}: it first drains every job in the critical queue and, only when that queue is empty, takes the job at the head of the priority queue. Draining the critical queue first ensures that jobs close to their deadlines are considered before higher-ranked jobs that can still wait. The decision logic and priorities used are discussed in \S~\ref{sec:scheduling}. Since every job is held in both the priority and the deadline queues, a job dispatched from one queue leaves a stale copy in the other. HyDRA records each dispatched job in an in-memory map; when the job selector later encounters a job already in the map, it discards that copy, removes the job from the map, and does not dispatch it again. Stale copies in the priority queue are also removed lazily during its periodic score updates.

A \textit{backfiller} starts the job at the head of the dispatch queue if the \textit{resource manager} reports enough free nodes. Otherwise, it backfills a later job that fits into the free nodes, provided this preserves the deadlines of waiting jobs and keeps the project within budget and the user's service deficit within a set tolerance. Each placed job is recorded in the \textit{scheduler plan} and appended to an FCFS \textit{executor queue}, which launches it on the HPC cluster. When a job starts or finishes, the resource manager updates the free nodes and charges the job's service to its user and project in the project budgets table, updating $S_{ij}(t)$, $S_j(t)$, and the deficits used for subsequent deadlines.

\section{Fairness-Aware Hierarchical Scheduling}\label{sec:scheduling}

In this section, we describe the mechanisms and scheduling logic of HyDRA.

\subsection{Hierarchical Entitlements and Priorities}\label{sec:entitlement}

Entitlement refers to the share of the cluster each project and each user should receive over time. The entitlement of a project is its weight divided by the total weight of all projects, $\hat{w}_j=w_j/\sum_{j'}w_{j'}$. For example, a project with weight 2, when the weights of all projects add up to 10, should receive 20\% of the cluster. Each project then shares its entitlement among its own users. At Level-2 and Level-3, each user receives a part of the project's entitlement in proportion to its weight. At Level-1, where the project does not assign weights to its users, all of its users receive equal shares ($E_{ij}= \dfrac{\hat{w}_j}{|\mathcal{U}_j|}$), where $|\mathcal{U}_j|$ is the number of users in project $p_j$. However, $E_{ij}=\hat{w}_j\,\dfrac{w_{ij}}{\sum_{i'}w_{i'j}}, \text{at Level-2 and Level-3}$. Job weights, which are used only at Level-3, do not change these entitlements; they only affect the priority of the job to which they belong.

Next, the priority $\rho_k$ of a job is obtained by multiplying the weights that apply at the chosen level. At Level-1, a job simply inherits the weight of its project, i.e., $\rho_k = w_j$. At Level-2, $\rho_k=w_jw_{ij}$, and at Level-3, $\rho_k=w_jw_{ij}w_k$. In this way, a job submitted by a high-priority user in a high-priority project receives a high priority, and vice versa.
The value of this product depends on which level is active and on the weight ranges the computing facility supports. For example, with weights between 1 and 5, the product range is$[1,5]$ at Level-1, but $[1,25]$ at Level-2. To maintain consistency across all the levels and across different facilities, HyDRA converts each priority to a fixed range of values $[1,5]$ before using it to compute a deadline. 
\begin{equation}
\tilde{\rho}_k=1+4\,\frac{\rho_k-\rho_{\min}}{\rho_{\max}-\rho_{\min}},
\label{eqn:priority-map}
\end{equation}
where $\rho_{\min}$ and $\rho_{\max}$ are the lowest and highest priorities that the chosen weights allow, and $\tilde{\rho}_k=3$ when all weights are equal. We avoid using ranges in [0,1], as smaller ranges in the denominator tend to significantly inflate the deadline. 

\subsection{Dynamic Priority Score}
The priority queue orders waiting jobs by a dynamic priority score that combines a job's priority, its age in the queue, and its size:
\begin{equation}
\pi_k(t)=\tilde{\rho_k}\,\frac{1+A_k(t)}{q_k^{\gamma}},\qquad A_k(t)=t-a_k,
\label{eqn:score}
\end{equation}
where $A_k(t)$ is the time the job has waited so far. $\rho_k$ places jobs in the queue according to their priority. The numerator grows the longer a job waits, so every waiting job slowly moves up the queue. The constant 1 in the numerator gives a newly arrived job a non-zero score. This dynamic priority score captures three behaviors. First, jobs of higher-priority projects and users rank higher. Second, jobs that have waited longer gradually rise in the queue, which prevents starvation. Third, smaller jobs rank higher since they are easier to place, with $\gamma=0$ ignoring job size and $\gamma=1$ fully normalizing by service demand. Since $\pi_k(t)$ grows linearly with $t$, two waiting jobs exchange rank at most once, and we characterize when this happens in \S~\ref{sec:analysis}. Aging alone, however, cannot bound how long a job waits. This is the role of the deadline, which we describe next.

\subsection{Adaptive Deadlines}
HyDRA assigns a deadline to each job during the initialization stage. The estimated deadline should reflect three factors: the size of the job, its importance, and the cluster's current workload. We denote the waiting-time bound of job $J_k$ as:
\begin{equation}
\Omega_k=\kappa\,\frac{q_k^{\gamma}}{\tilde{\rho}_k}\cdot[1+\,L(a_k)],\qquad D_k=a_k+\Omega_k.
\label{eqn:deadline}
\end{equation}
The first component is the \textit{base bound}. It grows with the service demand $q_k$ of the job, since a large job needs many nodes at once and is harder to place, and it shrinks with the importance of the job, $\tilde{\rho}_k$. The constant $\kappa>0$ scales all bounds to the cluster, and $\gamma$ is the same size exponent as in Eqn.~\ref{eqn:score}.
The second component adapts the bound to the cluster state upon job arrival. A bound computed from the job alone would promise the same wait during a burst as on an idle cluster, and would be missed if the queue is long. The \textit{backlog factor} $L(t)=Q(t)/\big(Q(t)+C(T)\big)$ measures how much work is waiting, i.e., the total service demand $Q(t)$ of queued jobs, relative to the node-time $C(T)=NT$ that the cluster can deliver within a window of backlog length $T$ (\S~\ref{sec:problem}). It ranges from 0 for an empty queue to nearly 1 for a long one.

For HyDRA, we design three deadline variants, one for each level of the hierarchy, which differ only in the priority $\tilde{\rho}_k$ used in Eqn.~\ref{eqn:deadline}. At Level-1, $\tilde{\rho}_k=\rho_k=w_j$, since project weights between 1 and 5 already lie in the target range. At Level-2 and Level-3, $\tilde{\rho}_k$ is obtained by converting $\rho_k=w_jw_{ij}$ and $\rho_k=w_jw_{ij}w_k$, respectively, to $[1,5]$ with Eqn.~\ref{eqn:priority-map}.

\subsection{Admission, Selection, and Fairness-Safe Backfilling}

\begin{algorithm}[t]
\footnotesize
\caption{HyDRA scheduling at a decision point $t$}
\label{alg:hydra}
\begin{algorithmic}[1]
\For{each job $J_k$ in the job buffer} \Comment{initializer}
  \State $q_k\gets n_k\tau_k$; compute $\rho_k$ and $D_k$ (Eqns.~\ref{eqn:priority-map},~\ref{eqn:deadline})
  \If{$q_k>B_j-S_j(t)$} \Comment{budget manager}
    \State reject $J_k$
  \Else
    \State insert $J_k$ into $\mathcal{Q}_\pi$ by $\pi_k(t)$ and into $\mathcal{Q}_D$ by $D_k$
  \EndIf
\EndFor
\State recompute $\pi_k(t)$ for all jobs in $\mathcal{Q}_\pi$; drop stale copies
\While{the head $J_h$ of $\mathcal{Q}_D$ has $D_h-t\le\delta$} \Comment{deadline handler}
  \State move $J_h$ to the critical queue $\mathcal{Q}_C$
\EndWhile
\State $\mathcal{B}\gets$ all jobs in $\mathcal{Q}_C$, or the head of $\mathcal{Q}_\pi$ if $\mathcal{Q}_C=\emptyset$ \Comment{job selector}
\For{each $J_k\in\mathcal{B}$}
  \If{$J_k\in\mathcal{M}$}
    \State remove $J_k$ from $\mathcal{M}$ \Comment{stale copy}
  \Else
    \State add $J_k$ to $\mathcal{M}$ and append it to the dispatch queue $\mathcal{Q}_S$
  \EndIf
\EndFor
\State $J_h\gets$ head of $\mathcal{Q}_S$ \Comment{backfiller}
\If{$n_h\le N_{\mathrm{free}}(t)$}
  \State \Call{Start}{$J_h$}
\Else
  \State $R_h\gets$ earliest time at which $n_h$ nodes are free
  \For{each later job $J_c\in\mathcal{Q}_S$ with $n_c\le N_{\mathrm{free}}(t)$}
    \If{$R'_h\le D_h$, $J_c$ is budget-feasible, and $F_{ij}(t)\ge-\zeta$}
      \State \Call{Start}{$J_c$} \Comment{$R'_h$: start of $J_h$ if $J_c$ runs}
    \EndIf
  \EndFor
\EndIf
\Procedure{Start}{$J_k$}
  \State record $J_k$ in the scheduler plan and append it to the executor queue
  \State update $N_{\mathrm{free}}(t)$, $S_{ij}(t)$, $S_j(t)$, and the deficits
\EndProcedure
\end{algorithmic}
\end{algorithm}

Algorithm~\ref{alg:hydra} summarizes how HyDRA combines these mechanisms at each scheduling decision point $t$. On arrival, the budget manager admits a job $J_k$ of project $p_j$ only if it is \textit{budget-feasible}, i.e., $q_k\le B_j-S_j(t)$, and otherwise rejects it early. Admitted jobs enter both the priority queue $\mathcal{Q}_\pi$, ordered by Eqn.~\ref{eqn:score}, and the deadline queue $\mathcal{Q}_D$, ordered by Eqn.~\ref{eqn:deadline}. The deadline handler moves every deadline-critical job into the critical queue $\mathcal{Q}_C$, and the job selector dispatches all of them before the head of $\mathcal{Q}_\pi$, using the map $\mathcal{M}$ to discard the stale copy of a job already dispatched from the other queue.

The backfiller starts the head $J_h$ of the dispatch queue $\mathcal{Q}_S$ if $n_h$ nodes are free. Otherwise, it computes the reservation $R_h$, the earliest time at which $n_h$ nodes become free based on the walltimes of running jobs, and scans the later jobs in $\mathcal{Q}_S$. A candidate $J_c$ of user $u_{ij}$ that fits in the free nodes starts only if two conditions hold: (i) $J_h$ can still start by its deadline, i.e., its estimated start $R'_h$ if $J_c$ runs satisfies $R'_h\le D_h$; and (ii) $J_c$ is budget-feasible. Condition (i) distinguishes HyDRA from EASY backfilling~\cite{Lifka1995EASY}, which forbids any delay to the reserved job: HyDRA lets a backfilled job delay the head job as long as the head job still meets its deadline, which raises utilization without breaking the waiting-time bound. Conditions (ii) prevent backfilling from exhausting a project's budget.

\subsection{Mathematical Analysis}\label{sec:analysis}
Since the priority queue always serves jobs in order of their priority score, the way $\pi_k(t)$ changes over time decides when one waiting job overtakes another. We therefore analyze how the score of a job evolves while it waits in the queue, to show that aging reorders the jobs in priority queue in a predictable way.

Substituting the waiting time $A_k(t)=t-a_k$ into Eqn.~\ref{eqn:score} shows that the score of a waiting job is a straight line in time:
\begin{equation}
\pi_k(t)=b_k+m_k\,t,\qquad m_k=\frac{\tilde{\rho_k}}{q_k^{\gamma}},\qquad b_k=m_k\,(1-a_k).
\label{eqn:score-line}
\end{equation}
The slope $m_k$ sets how fast a job gains priority while it waits. It is larger for jobs with a higher priority $\tilde{\rho_k}$, and smaller for larger jobs when $\gamma>0$, so a large job of a low-priority user rises the slowest. The intercept $b_k$ also depends on the arrival time, so a job that arrives later starts lower. Two jobs with the same priority and size have parallel lines that never cross, so they are served in the order in which they arrived.

Since each score is a line, two waiting jobs $J_a$ and $J_b$ with different slopes swap places at most once, at the time their lines cross:
\begin{equation}
t^*=\frac{b_b-b_a}{m_a-m_b},\qquad m_a\neq m_b.
\label{eqn:crossing}
\end{equation}
The swap takes effect only if $t^*$ falls after both jobs have arrived and before either of them starts. After that, the job with the larger slope stays ahead for as long as both jobs wait. This analysis applies only while a job is in the queue, i.e., from $a_k$ until it is sent for execution. 

This analysis helps strengthen the fact that aging in HyDRA never makes the jobs oscillate inside the queue. Every pair of jobs swaps places at most once, and a job with a larger slope $m_k$ eventually overtakes any job with a smaller slope that is still waiting.

\section{Evaluation}\label{sec:evaluation}

\subsection{Implementation and Setup}  

We implement HyDRA in C++20. Its components, from the job buffer to the executor queue, run as they would in a deployment, while the cluster and its resource manager are simulated: the executor holds $n_k$ nodes of the simulated cluster for the trace runtime $r_k$ of each job. The baselines share this code and replace only the ordering of the priority queue, except PCS and ANL, which run in a standalone multi-queue simulator written in C++17. Each run executes on one NVIDIA DGX Spark node (20-core ARM CPU, 128\,GB of memory) of a university HPC cluster.

\subsection{Baselines}
We compare HyDRA against six baselines: three conventional scheduling policies, two SOTA fair schedulers that we adapt to rigid HPC jobs, and the production policies of Argonne National Laboratory (ANL). PCS and ANL run in a standalone multi-queue simulator, which we implement on top of HyDRA. All other baselines run within our architecture (Section~\ref{sec:architecture}), with the deadline queue disabled and the dynamic priority score $\pi_k(t)$ replaced by the ordering key of the baseline.

\textit{Shortest Job First (SJF)} orders jobs by their requested walltime $\tau_k$. Because short jobs are expected to release their resources quickly, SJF tends to reduce average wait time. However, it can starve long-running jobs, and it ignores job width and project membership entirely. \textit{Priority} orders jobs by the priority $w_j$ of their project, so it honors the relative importance of projects but not how much service they have already received. \textit{Weighted Fair Queueing (WFQ)}~\cite{Parekh1993GPS} orders jobs by a virtual finish tag computed from their service demand and the weight of their project, so that each project is served in proportion to its weight.

\textit{Themis}~\cite{Mahajan2020Themis} introduced finish-time fairness for GPU clusters. Our Themis-inspired variant keeps its core idea and prioritizes the jobs of the project with the largest service deficit over a rolling $300$~s window, but omits the leases and auctions of the original. \textit{PCS}~\cite{Faisal2024PCS} balances performance, fairness, and predictability by sorting jobs into size classes, which we implement as four queues split by requested walltime:  A: $\leq300\,s$; B: ($>300$, $\leq1,800\,s$); C: ($>1,800\,s$, $\leq 7,200\,s$); D: $>7,200\,s$. Jobs are served FIFO within a queue, and queue $i$ receives a share of the resources weighted by $e^{-\omega i}$, so that short jobs are preferred without starving long ones.

\textit{ANL} models the job-sort formula of the PBS scheduler at
ALCF~\cite{openpbs2023jobsort,argonne_pbsq}. It ranks each waiting job $k$ of
project $j$ by
\begin{equation}
S_k(t) = 51 + 10^{5}\,\frac{e_k(t)^{2}}{\tilde{w}_k^{3}}\,\pi_j\,\frac{n_k}{N}
       + \frac{e_k(t)}{1800},
\label{eq:anl}
\end{equation}
where $e_k(t)$ is the job's waiting time, $\tilde{w}_k$ is its requested
walltime clamped to $[6, 12]$\,hours, $\pi_j$ is the project priority, and
$n_k/N$ is the fraction of the system the job requests.

To ensure a fair comparison, all baselines run the same workloads on the same simulated cluster as HyDRA, use the same reservation-based backfilling, and see only the requested walltimes, never the actual runtimes.

\subsection{Workloads}
\begin{table}[t]
\centering
\footnotesize
\caption{Production traces used for synthetic workload generation.}
\label{tab:workloads}
\setlength{\tabcolsep}{1pt}
\begin{tabularx}{\columnwidth}{@{}>{\raggedright\arraybackslash\hsize=1.2\hsize}X>{\raggedright\arraybackslash\hsize=0.8\hsize}X@{}}
\toprule
Source trace & \textbf{Polaris (ALCF)} \\
\midrule
Year, duration & 2025, 365\,d  \\
Nodes $\times$ cores & $560\times64$  \\
Projects, users & 304, 953 \\
Jobs & 208,405 \\
Active users per project\textsuperscript{a} & 1.17 / 1 / 55  \\
Peak active projects / users\textsuperscript{b} & 51 / 100  \\
\bottomrule
\end{tabularx}
\\[2pt]
{\raggedright \textsuperscript{a} Mean / median / maximum over active projects. \textsuperscript{b} In any 2\,h window. }
\end{table}

For our experiments, we generate workloads by sampling projects, users, and jobs from the production traces of the 2025 job trace of ALCF Polaris~\cite{alcf_cpu_dataset}. The characteristics are summarized in Table~\ref{tab:workloads}. The trace is used without postprocessing, as each of its fields maps directly onto the job model of Section~\ref{sec:problem}. Moreover, we observed that the runtime for a job exceeded their walltimes in $19\%$ of the cases, with a p99 overshoot of $0.1\%$, so we let the jobs continue to execute even if the runtime exceeds their walltime. 

For our experiments, we generate candidate windows of durations of 
2, 6 and 12 hours with each window starting 1 hour apart. For each window, we calculate the load as number of active users for that window. Based on the quantile value (l) of the load distribution in that window, we categorize the windows as unconditioned (l=0), busy (l=0.9), bursty (l=0.95) and highly bursty (l=0.99). Based on the kind of workload required for each window, we then start by randomly picking one eligible window of the required category. 
The number of active project in that window is extracted, along with its associated users and jobs. If the number of projects are fewer than what's specified in the workload, we append projects from other eligible windows of the same category. If a chosen window holds more projects than requested, we sample the required number of projects from it without replacement and discard the rest. The arrival time of jobs in those windows are taken relative to the start of the window. Project, user and job priorities are drawn uniformly from $\{1,\dots,5\}$, where $5$ is the highest priority. Maximum possible deliverable service in a window is the total capacity $N \times $ the budget window $T=300s = C(T)$. If there is a job with $q_k > C(T)$, it would get rejected by the budget manager. For our experiments, we create $3$ workloads for each duration: Small, Medium and Large with $50$, $100$ and $1000$ projects respectively.  

Overall, our workload spans $197-12,356$ jobs with $54-4,365$ users. Unless stated otherwise, all HyDRA variants use the same hyperparameters on every workload. We set $\kappa=5$, $\gamma=0.75$, with waiting times in seconds, after sweeping each parameter on its own while keeping the others at their defaults, over $\kappa\in\{1,5,10,100\}$, $\gamma\in\{0.5,0.75,1\}$. The backlog factor uses a window of $T=30$\,s, a job becomes deadline-critical when it is within $\delta=300$\,s of its deadline, the deficits use $\epsilon=10^{-12}$.

\subsection{Metrics}

We evaluate HyDRA on three aspects: fairness, deadline, and system performance metrics.

Fairness is measured by comparing the service received against entitlement. If the service $S(t)$ delivered up to time $t$ were divided exactly according to entitlement (Section~\ref{sec:entitlement}), project $p_j$ and user $u_{ij}$ would have received $S_j^*(t)=\hat{w}_jS(t)$ and $S_{ij}^*(t)=E_{ij}S(t)$, respectively. The \textit{service lag} is the gap between this entitled service and the service actually received:
\begin{equation}
\Delta_j(t)=S_j^*(t)-S_j(t),\qquad \Delta_{ij}(t)=S_{ij}^*(t)-S_{ij}(t).
\label{eqn:lag}
\end{equation}
A positive lag means that the project or user has received less than its entitlement and is under-served, while a negative lag means that it is over-served. The user lag is measured end to end, i.e., against the user's share of the whole cluster rather than of its project, so it captures unfairness at both levels of the hierarchy.
To summarize fairness across all projects in a single number, we use Jain's fairness index~\cite{Jain1984Fairness} over the ratios $x_j=S_j(t)/S_j^*(t)$ of received to entitled service:
$\mathrm{JFI}=\frac{\big(\sum_j x_j\big)^2}{|P|\sum_j x_j^2}$, where $|P|$ is the number of projects. The index equals 1 when every project receives exactly its entitlement, and it decreases as service becomes more uneven. We compute the same index end to end over the users of all projects, using the ratios $x_{ij}=S_{ij}(t)/S_{ij}^*(t)$ and the total number of users in place of $|P|$. Both indices include only the projects and users with a positive entitlement.

Further, we evaluate how close the estimated waiting-time bounds are. The \textit{on-time rate} is the fraction of jobs that start by their deadline, i.e., with $s_k\le D_k$. A high on-time rate alone, however, can be achieved with very loose deadlines that tell users little. We therefore also report the \textit{deadline deviation} $s_k-D_k=\omega_k-\Omega_k$, the difference between the actual and the promised wait. A negative deviation is slack that the job did not need, while a positive deviation means that HyDRA over-promised, so an accurate estimate keeps the deviation close to zero and rarely positive. Since the baselines do not assign deadlines, we evaluate them against the deadlines that HyDRA assigns to the same jobs at Level-1.

To compare waits across jobs of different sizes, we report the \textit{bounded slowdown}~\cite{Feitelson1997Theory} of each job, i.e., its waiting time $\omega_k$ relative to its runtime $r_k$:
$\mathrm{BSD}_k=1+\frac{\omega_k}{\max(r_k,\tau_{\min})}$, where the threshold $\tau_{\min}=10$\,s keeps very short jobs from inflating the metric. A job that starts immediately has $\mathrm{BSD}_k=1$, and a job that waits as long as it runs has $\mathrm{BSD}_k=2$. To check that HyDRA does not shorten the waits of high-priority jobs at the expense of low-priority ones, we report the median and 95th percentile of $\mathrm{BSD}_k$ separately for each project-priority and user-priority tier.

Finally, we report the average and 95th-percentile queue wait $\omega_k$, and the \textit{cluster utilization} $S(e_{\max})/C(e_{\max})$, i.e., the fraction of the node-time that the cluster could deliver until the last job completes at $e_{\max}=\max_k e_k$ that is used by running jobs. These show what HyDRA's fairness and deadlines cost in performance.

\subsection{Results} 

\subsubsection{Comparison with Baselines}

\begin{figure}[t]
    \centering
    \subfloat[On-time rate]{
    \includegraphics[width=0.45\linewidth]{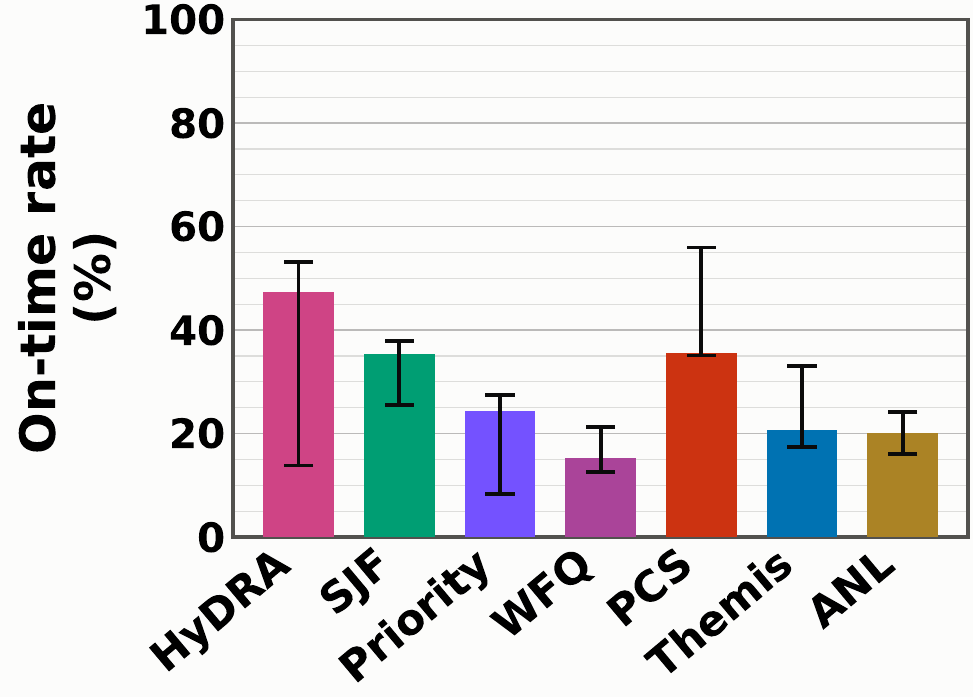}\label{fig:baseline-large-ontime}}
    \subfloat[Project Jain's index]{
    \includegraphics[width=0.45\linewidth]{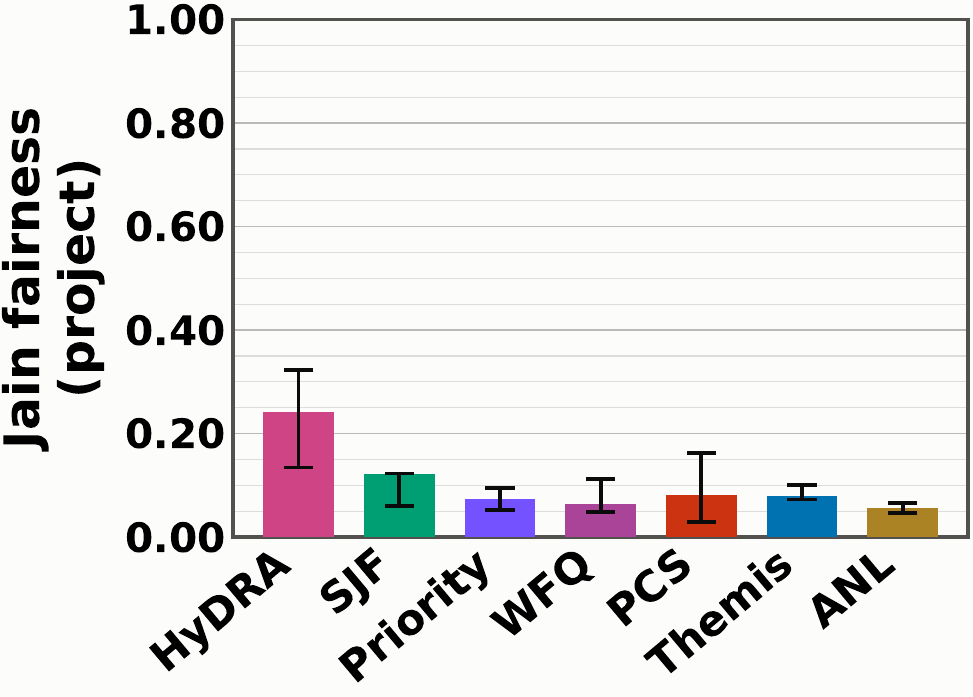}\label{fig:baseline-large-jfi}}\\
    \subfloat[Project service lag (p95)]{
    \includegraphics[width=0.45\linewidth]{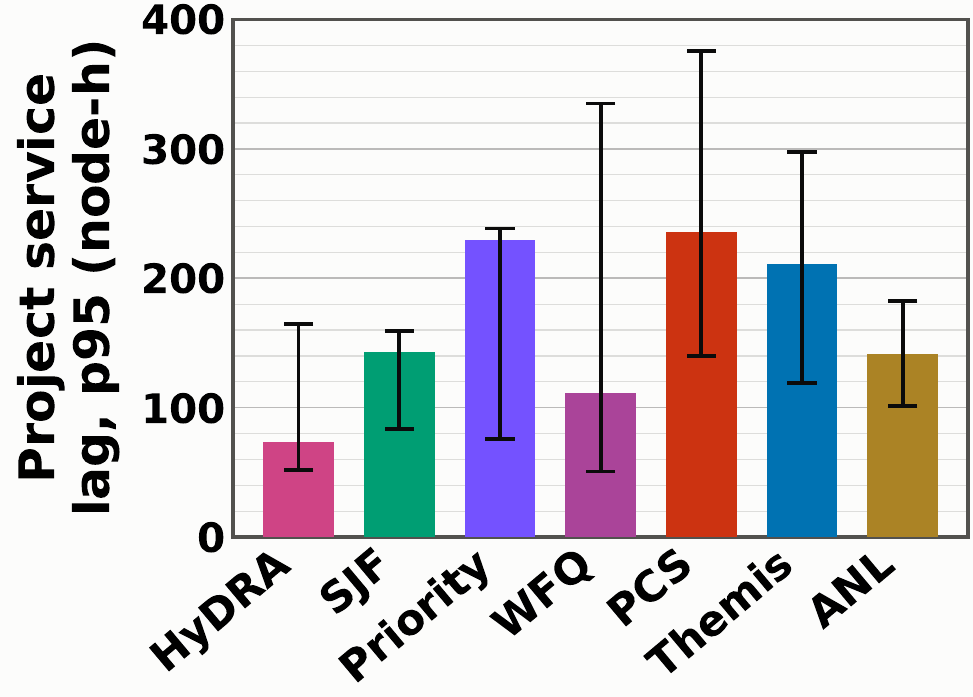}\label{fig:baseline-large-lag}}
    \subfloat[Bounded slowdown (p95). Log scale.]{
    \includegraphics[width=0.45\linewidth]{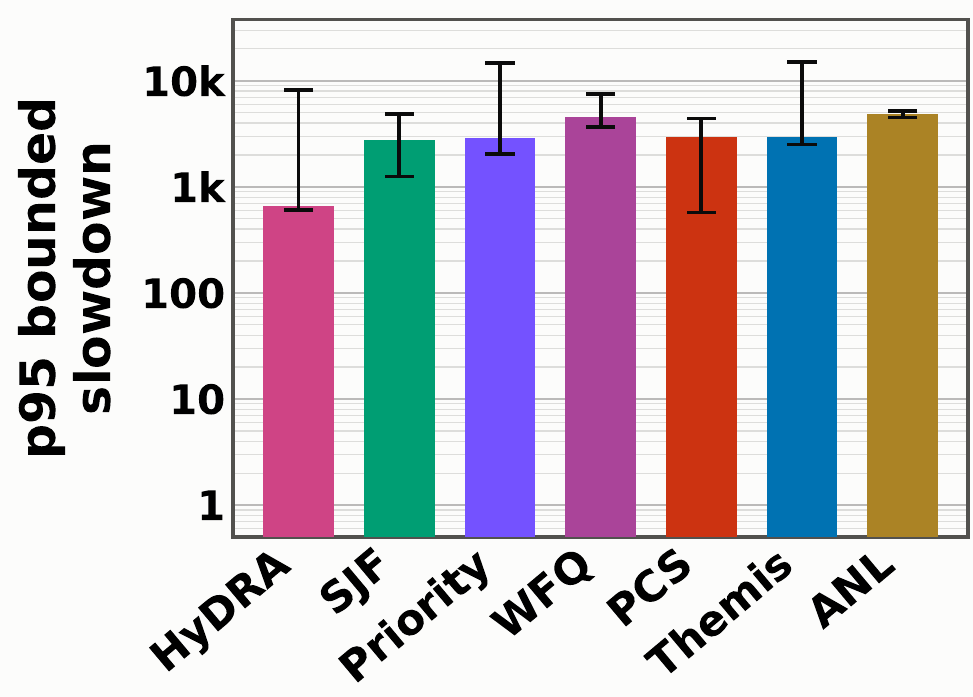}\label{fig:baseline-large-bsd}}
    \caption{Comparison across the Large workloads. }
    \label{fig:baseline-large}
\end{figure}

We compare HyDRA with the baselines on the Polaris workloads of the unconditioned load, for the three periods (2, 6, and 12\,h) and three workload sizes. We run HyDRA at Level-1 in this comparison. Fig.~\ref{fig:baseline-large} shows the on-time rate, the project-level Jain's index, the 95th-percentile project service lag, and the 95th-percentile bounded slowdown of each scheduler. Each bar represents the median across the three periods, and the whiskers indicate the minimum and maximum. Every scheduler is compared against the deadlines that HyDRA assigns to the same jobs. ANL did not finish the 12\,h Large workload within $12$ hours, so we mark it as DNF and do not include it in the bars. 

\begin{figure}[t]
    \centering
    \includegraphics[width=0.9\linewidth]{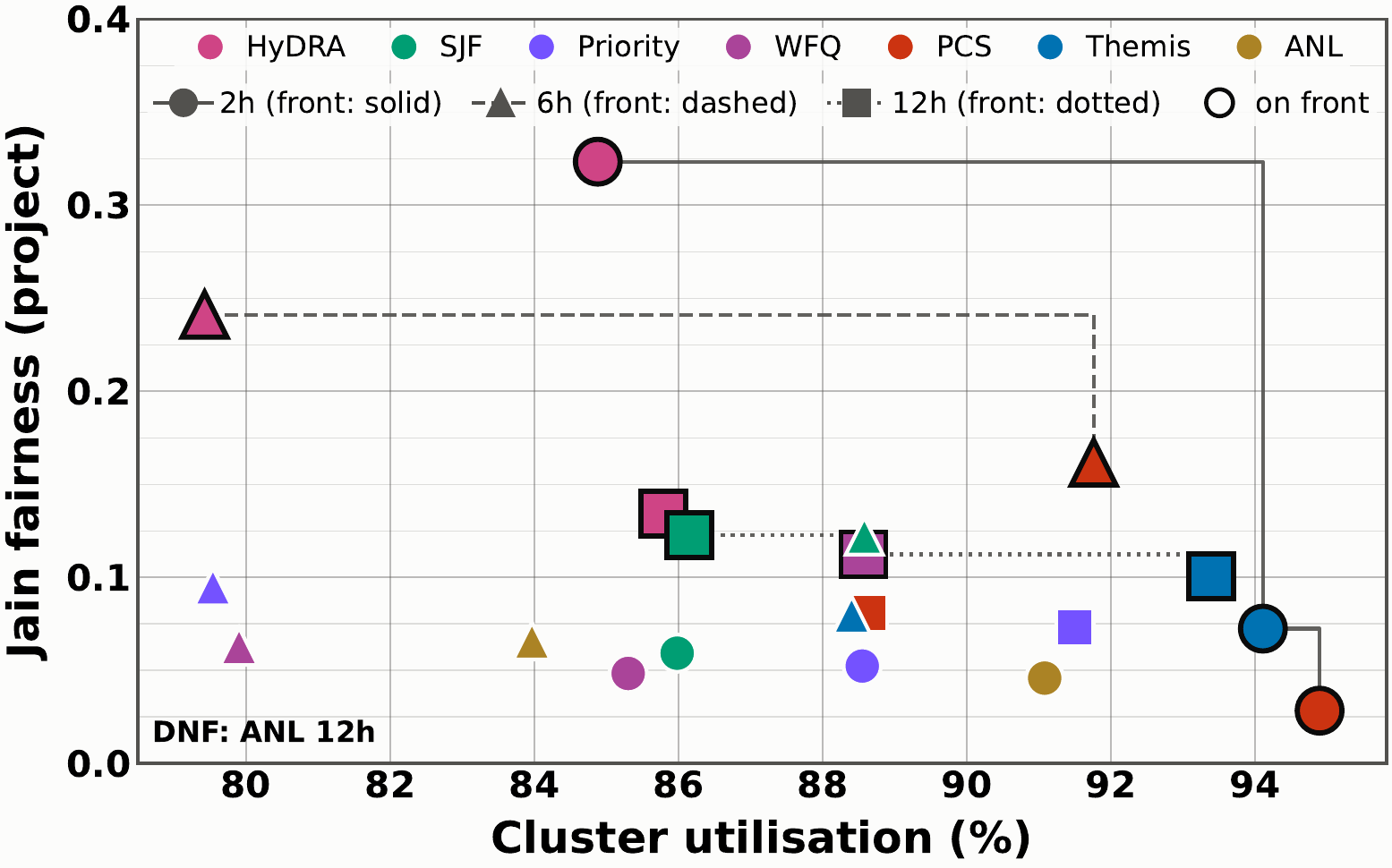}
    \caption{Cluster utilization against project-level Jain's index on the Large workloads. Marker shape shows the duratin of experiment, and the step lines connect the Pareto front of each period, whose members are outlined.}
    \label{fig:pareto}
\end{figure}

\textit{\uline{Light load.}} The Small and Medium workloads keep the utilization below 43\%, so jobs rarely wait. Even in the worst case, every scheduler starts all jobs on time with a Jain's index of at least 0.97 (Table~\ref{tab:baseline-sm}). Without contention, the ordering policy hardly matters, so we focus on the Large workloads for the rest of this section.

\textit{\uline{Fairness.}} Under heavy load, HyDRA is clearly the fairest scheduler (Fig.~\ref{fig:baseline-large-jfi}). Its median project-level Jain's index is 0.24, twice that of SJF (0.12), the strongest baseline, and upto $4\times$ compared to others. The absolute values are low for all schedulers because $1,000$ projects share $560$ nodes over a few hours. To put them in perspective, the index equals $k/n$ when only $k$ of $n$ projects share the service evenly and the rest receive none, so HyDRA serves nearly $25\%$ projects evenly, whereas the baselines serve only 6--12\%. Interestingly, Themis and WFQ are no fairer than SJF, even though they explicitly balance service between projects. Balancing tags or deficits does not by itself start the jobs of an under-served project when they need more nodes than are free, whereas HyDRA's deadlines cap how long any job can be passed over.

\begin{table}[t]
\centering
\footnotesize
\caption{Worst value of each metric over all schedulers and over the 2, 6, and 12\,h Small and Medium workloads, and the scheduler with that value.}
\label{tab:baseline-sm}
\setlength{\tabcolsep}{3pt}
\begin{tabularx}{\columnwidth}{@{}l*{5}{>{\centering\arraybackslash}X}@{}}
\toprule
 & \textbf{On-time} & \textbf{Jain's} & \textbf{Lag p95} & \textbf{BSD} & \textbf{Utilization} \\
 & \textbf{(\%)} & \textbf{index} & \textbf{(node-h)} & \textbf{p95} & \textbf{(\%)} \\
\midrule
Worst value & 100 & 0.97 & 8.38 & 2.23 & 4.5--42.7 \\
Scheduler & All & Themis & Priority & Priority & All \\
\bottomrule
\end{tabularx}
\end{table}

\textit{\uline{Waiting-time guarantees.}} HyDRA maintains a high on-time completion rate, a median of $47\%$ against at most $36\%$ for the baselines (Fig.~\ref{fig:baseline-large-ontime}), and its tail slowdown is 4.2$\times$ lower than that of the next best baseline (Fig.~\ref{fig:baseline-large-bsd}). 
This advantage shrinks when the overload lasts 12\,h, since the backlog term can at most double a bound and most queued jobs then become deadline-critical.

\textit{\uline{Service lag.}} The 95th-\%ile project lag in Fig.~\ref{fig:baseline-large-lag} shows how far the hardest-hit projects fall behind their entitlement. HyDRA keeps it at a median of 74 node-hours, 34\% below WFQ (111), the closest baseline, and about half that of SJF and ANL (143 and 142). Themis, Priority, and PCS leave the worst-off projects 211--235 node-hours behind, about three times as far as HyDRA. WFQ also has the highest variation, so how well it protects lagging projects depends strongly on the workload.

\begin{figure*}[t]
\centering
\subfloat[Level-1 ]{%
  \includegraphics[width=0.6\columnwidth]{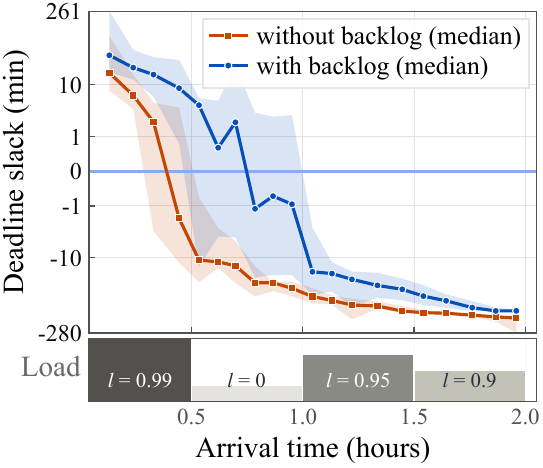}%
  \label{fig:level-1-bursts}}
\subfloat[Level-2 ]{%
  \includegraphics[width=0.6\columnwidth]{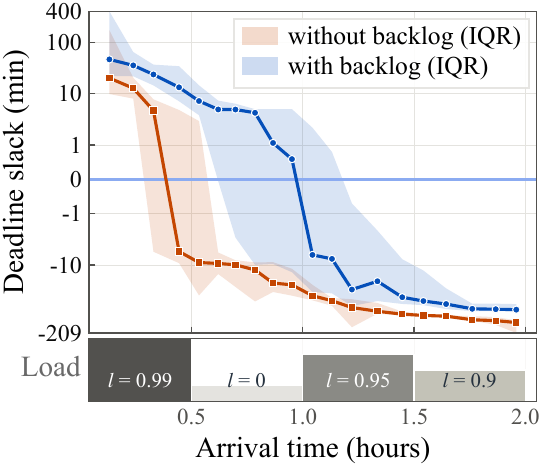}%
  \label{fig:level-2-bursts}} 
\subfloat[Level-3 ]{%
  \includegraphics[width=0.6\columnwidth]{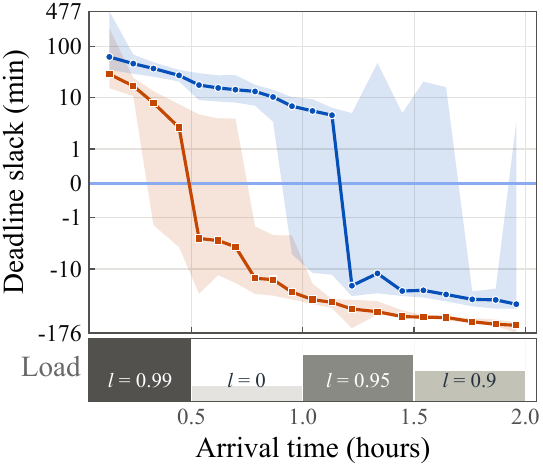}%
  \label{fig:level-3-bursts}}
\caption{HyDRA adapts better to bursty workload when adapting to the cluster current available capacity. Plots show the median and inter-quartile range (IQR) of Hydra variants.}
\label{fig:deadline-vs-bursts}
\end{figure*}

\textit{\uline{Fairness versus utilization.}} HyDRA's fairness comes at a moderate cost in utilization. Fig.~\ref{fig:pareto} plots the cluster utilization against the project-level Jain's index for the Large workloads. As we observe, HyDRA has the lowest utilization for every period. This is expected, since HyDRA schedules jobs by their deadlines and their projects' entitlements rather than by how easily they fit. A wide job close to its deadline, for example, holds nodes idle until enough of them are free, whereas PCS prefers short jobs that keep filling the cluster. Nevertheless, HyDRA lies on the Pareto front for every period, at its high-fairness end, and the baselines with the highest utilization are 11$\times$, 1.5$\times$, and 1.3$\times$ less fair at 2, 6, and 12\,h, respectively.
Overall, HyDRA achieves the best balance between fairness and waiting-time guarantees under heavy load.

\subsubsection{Performance of HyDRA in presence of bursts} 

Next, to understand the effect of the backlog term, we compare the performance across all the levels of hierarchy. Since the main focus of the experiment is to observe the reaction to a sudden change in workload activity, we evaluate them for a mixed workload. Here, the load distribution is shuffled randomly every $30$ minutes between normal (l=0), busy (l=0.9), bursty (l=0.95) and highly bursty (l=0.99) for a 2 hour experiment.

Fig.~\ref{fig:deadline-vs-bursts} shows how HyDRA reacts to this workload. The very first window is highly bursty, so the queue fills up quickly. Without the backlog term, HyDRA struggles from the start: its deadlines depend only on the job itself, so it keeps promising the same waits even as the queue grows. With the backlog term, the deadlines grow with the queued demand $Q(t)$, since $L(t)=Q(t)/\big(Q(t)+C(T)\big)$ rises almost linearly with $Q(t)$ while the queue is small compared with the node-time $C(T)$ that the cluster can deliver. This keeps the promised waits realistic for a long time. Once the queue becomes much larger than the capacity, i.e., $Q(t)\gg C(T)$, $L(t)$ approaches 1 and the deadlines stop growing with the queued demand.

\begin{figure}[t]
    \centering
    \subfloat[Polaris queue distribution]{
    \includegraphics[width=0.47\linewidth]{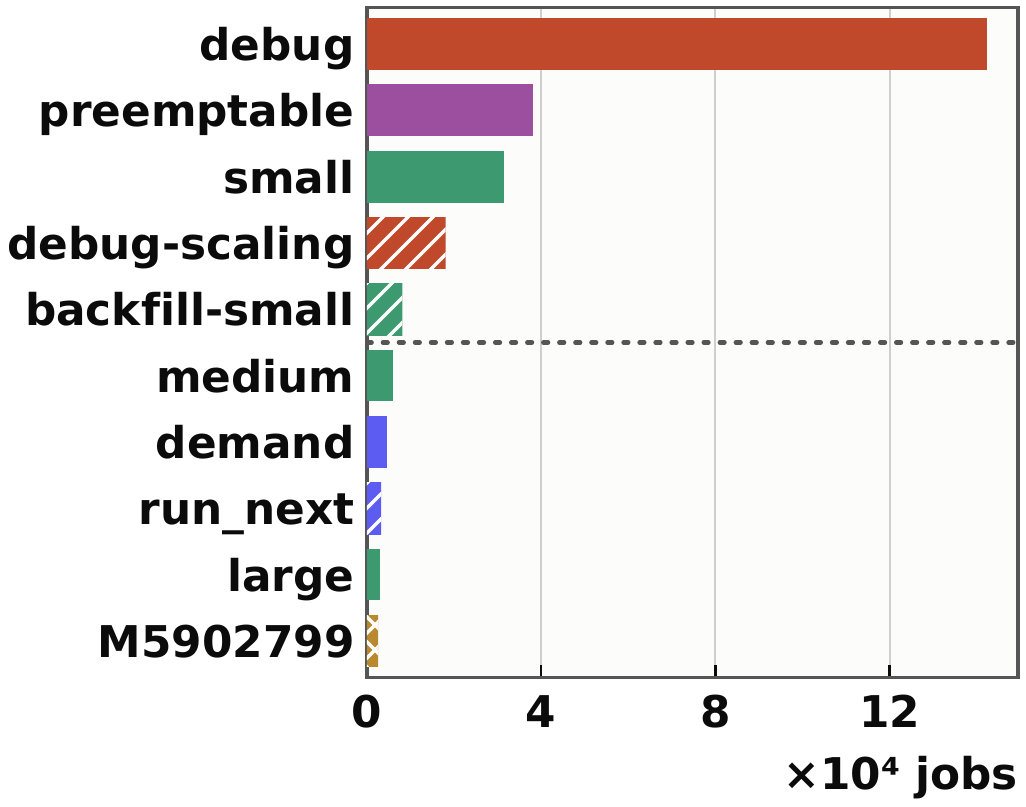}\label{fig:queue-dist}}
    \subfloat[Effect of the backlog term]{
    \includegraphics[width=0.52\linewidth]{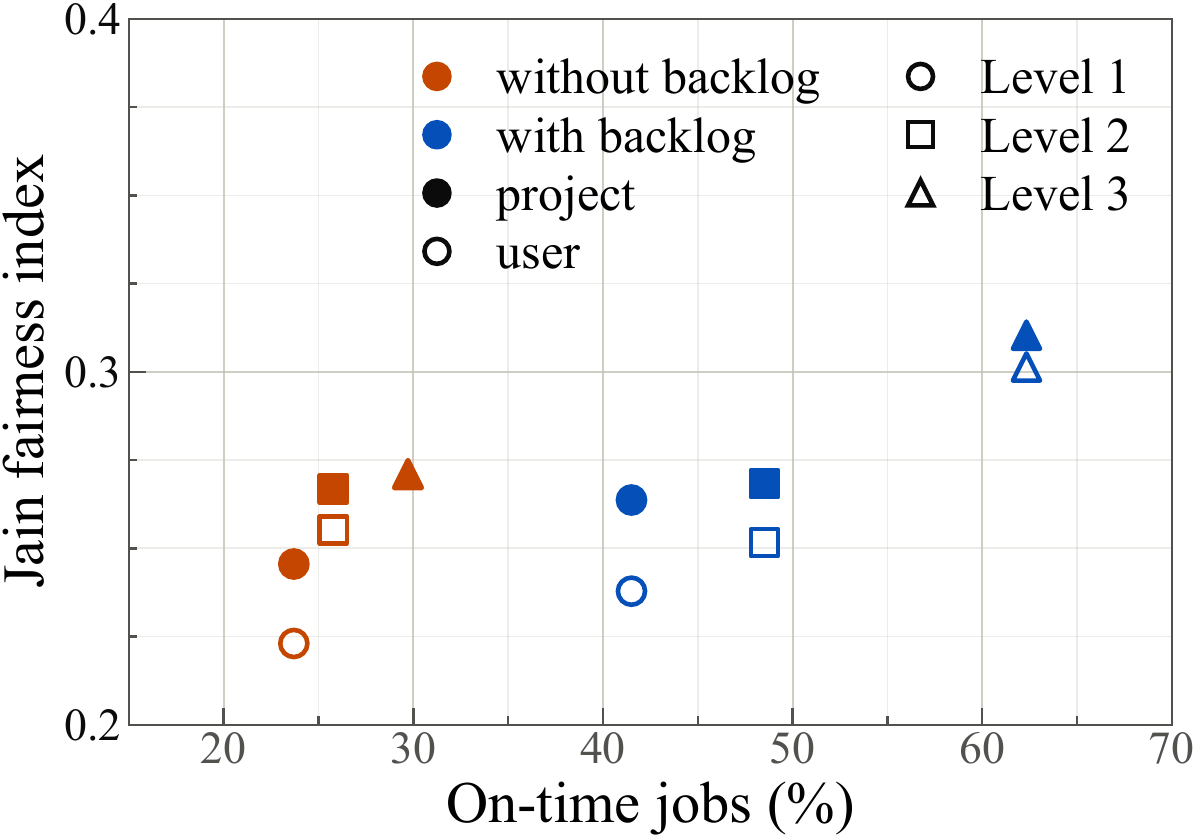}\label{fig:fairness-vs-bursts}}
    \caption{(a) Share of jobs in the ten most used Polaris queues, from which the queue-based job priorities $w_k$ are sampled (Table~\ref{tab:queue-priority}). (b) Jain's index and on-time rate of HyDRA at each hierarchy level, with and without the backlog term, on the mixed workload.}
    \label{fig:queues-and-backlog}
\end{figure}

Because the backlog term keeps deadlines achievable for longer, jobs also maintain a positive slack, i.e., time left before their deadlines, which raises the on-time rate (Fig.~\ref{fig:fairness-vs-bursts}). It improves fairness as well. While the slack is positive, few jobs are deadline-critical, so most jobs are dispatched from the priority queue, which ranks them by their priority, waiting time, and size. Once the slack becomes negative, more and more jobs become deadline-critical and are dispatched from the deadline queue instead, which serves them only in the order of their deadlines. Hence, the longer a scheduler stays in the positive-slack region, the fairer it is.

\subsubsection{Performance of HyDRA across hierarchy levels}

HyDRA supports hierarchical priorities across three entities: projects, users, and jobs. The priority $\rho_k$ of a job, and hence its deadline $D_k$, combines the weights that apply at each of the three levels (Section~\ref{sec:entitlement}). In this experiment, we study how the distribution of these priorities affects the performance of HyDRA.
During workload generation, the project, user, and job priorities $w_j$, $w_{ij}$, and $w_k$ are each sampled uniformly from $\{1,\dots,5\}$. Their products, however, are not uniform: the hierarchical priorities $w_j$, $w_jw_{ij}$, and $w_jw_{ij}w_k$  increasingly skew towards low values from Level-1 to Level-3.

\begin{table}[t]
  \centering
  \footnotesize
  \caption{Job priority assigned to each of the five most frequent Polaris queues.
    Shares are each queue's fraction of the jobs in these five queues available in the traces.}
  \label{tab:queue-priority}
  \begin{tabularx}{\columnwidth}{@{}c l r X@{}}
    \toprule
    Priority & Queue & Share & Basis \\
    \midrule
    5 & \texttt{debug}          & 59.74\% & Short jobs, dedicated nodes \\
    4 & \texttt{debug-scaling}  &  7.59\% & Above regular production \\
    3 & \texttt{small}          & 13.23\% & Normal production queue \\
    2 & \texttt{preemptable}    & 16.01\% & Documented as below demand and can be terminated \\
    1 & \texttt{backfill-small} &  3.43\% & Documented as low priority \\
    \bottomrule
  \end{tabularx}
\end{table}

On Polaris, jobs are submitted to dedicated queues according to their role (e.g., \texttt{debug} or \texttt{backfill}) or their size (e.g., \texttt{small} or \texttt{large}). Since a job's queue already reflects its role and size, it translates directly into a job priority. We therefore also generate another workload by sampling the job priority $w_k$ from the distribution of jobs across these queues, instead of uniformly. Based on Fig.~\ref{fig:queue-dist}, we map each queue to a priority in $\{1,\dots,5\}$, as listed in Table~\ref{tab:queue-priority}.

We compare the deadlines of four variants: Level-1, Level-2, Level-3, and Level-3 with queue-based job priorities. With the default sampling, $w_j$, $w_{ij}$, and $w_k$ are drawn independently and uniformly from $\{1,\dots,5\}$, so each has an expected value of 3. By independence, the expected priority is therefore $E[w_j]=3$ at Level-1, $E[w_jw_{ij}]=E[w_j]\,E[w_{ij}]=9$ at Level-2, and $E[w_jw_{ij}w_k]=E[w_j]\,E[w_{ij}]\,E[w_k]=27$ at Level-3. Now, the expectation for capped priority $\tilde{\rho_k}$ for each level is: 
\begin{align*}
\intertext{Level 1 ($\rho_{\min}=1,\ \rho_{\max}=5$), $\tilde\rho_k = w_j$, so}
  E[\tilde{\rho}_k] &= E[w_j] = 3
\intertext{Level 2 ($\rho_{\min}=1,\ \rho_{\max}=25$),}
  E[\tilde{\rho}_k] &= 1 + 4\,\frac{E[w_j w_{ij}] - 1}{24}
                     = 1 + 4\cdot\frac{9-1}{24} = \frac{7}{3} \approx 2.33
\intertext{Level 3 ($\rho_{\min}=1,\ \rho_{\max}=125$),}
  E[\tilde{\rho}_k] &= 1 + 4\,\frac{E[w_j w_{ij} w_k] - 1}{124}
                     = 1 + 4\cdot\frac{27-1}{124} = \frac{57}{31} \approx 1.84
\end{align*}

\begin{figure}[t]
    \centering
    \subfloat[All levels (1, 2, 3)]{
    \includegraphics[width=0.49\linewidth]{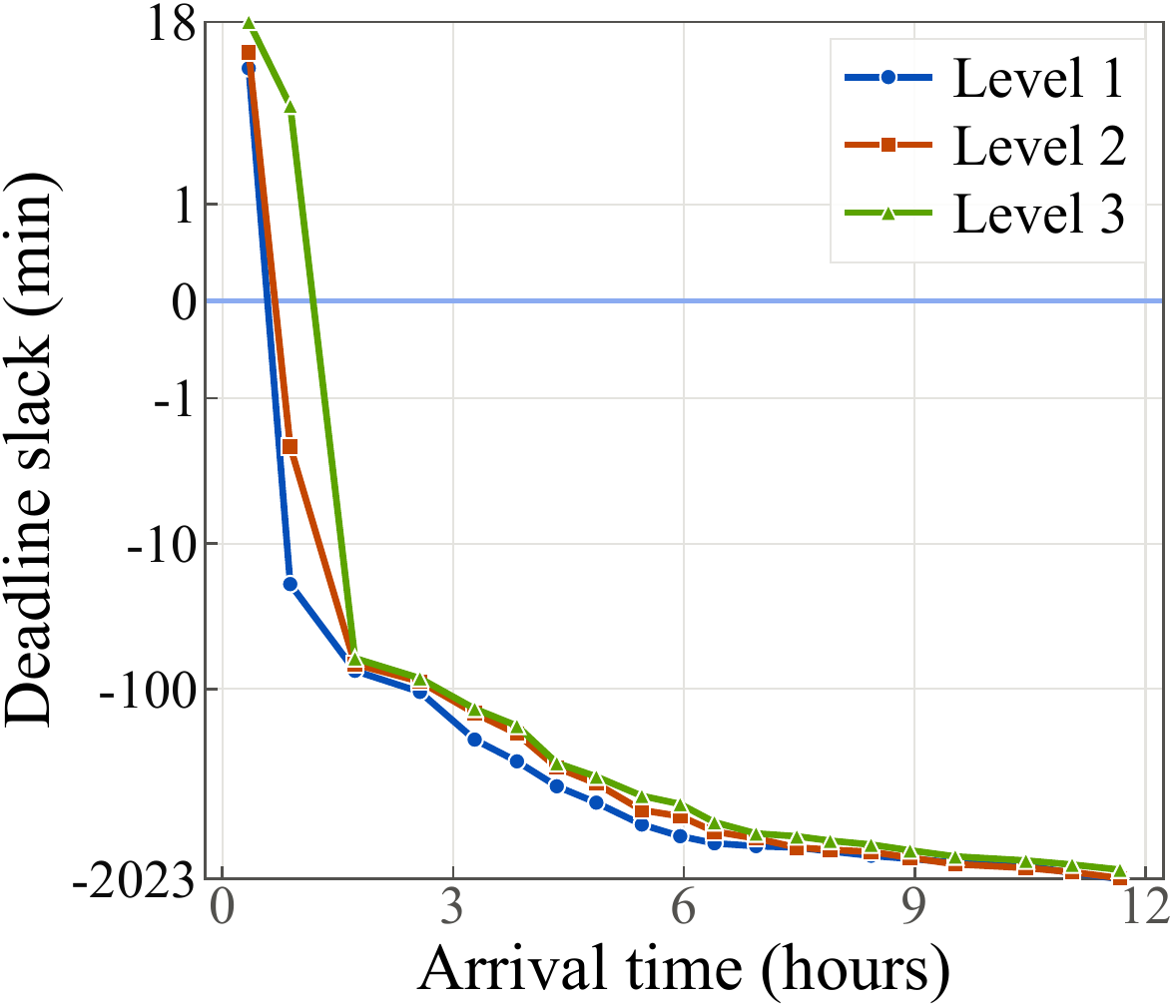}
    \label{fig:deadline-all-levels}}
    \subfloat[Level 3 (Uniform and Queue sampling)]{
    \includegraphics[width=0.49\linewidth]{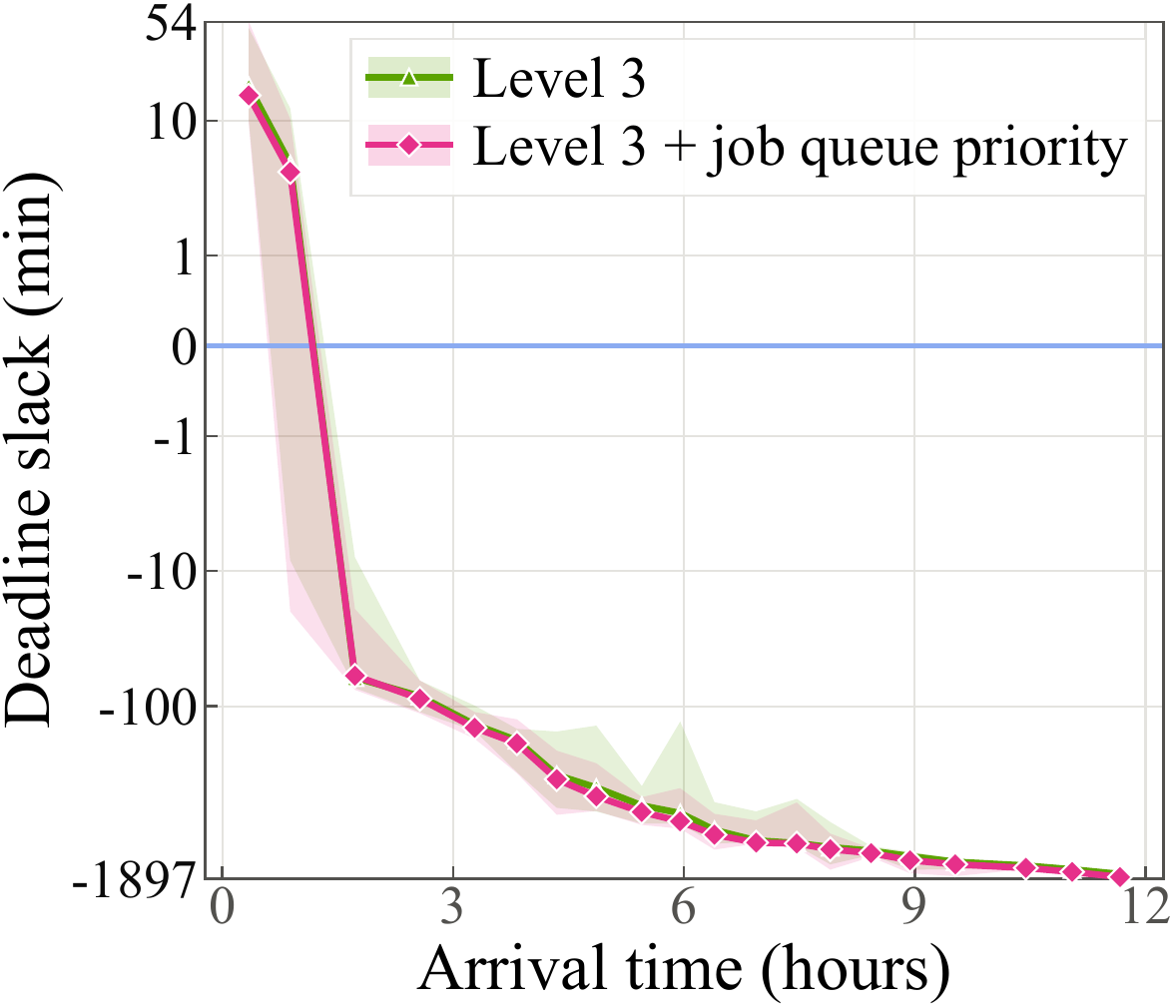}
    \label{fig:deadline-level3-uniform}}\\
     \subfloat[Level 3 (Uniform and Queue sampling)
     ]{
    \includegraphics[width=0.55\linewidth]{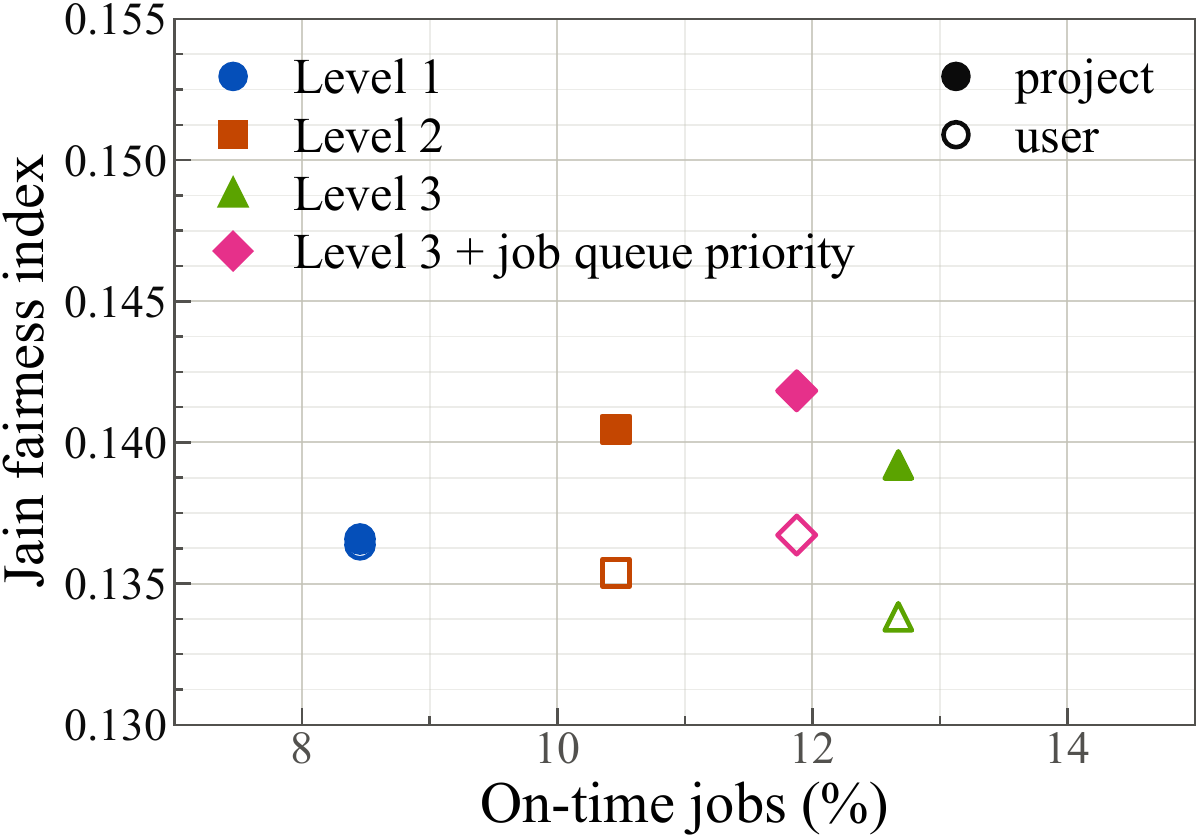}
    \label{fig:deadline-level3-trace}}
    \caption{Deadline slack vs. arrival for all hierarchical variants on Polaris (12 hr)}
    \label{fig:deadline-arrival-levels}
\end{figure}

As a result, $E[\tilde{\rho}_k]$ decreases as the hierarchy level increases. Since the waiting-time bound $\Omega_k$ is inversely proportional to $\tilde{\rho}_k$, the deadlines grow with the level: Fig.~\ref{fig:deadline-all-levels} shows that Level-3 deadlines exceed those of Level-2, which in turn exceed those of Level-1, for most of the run. Queue-based job priorities have the opposite effect. Under the default Level-3, $w_k$ is uniform, whereas with job-queue priorities its distribution is heavily skewed toward high weights (Fig.~\ref{fig:queue-dist}). Since $\rho_k=w_jw_{ij}w_k$ with independent factors, a larger $E[w_k]$ raises $E[\rho_k]$ and, because the conversion to $[1,5]$ is linear, also $E[\tilde{\rho}_k]$. Jobs under Level-3 with job-queue priorities therefore receive tighter deadlines on average (Fig.~\ref{fig:deadline-level3-uniform}).

\subsubsection{Performance of HyDRA against varying demand conditions}

\begin{figure}[t]
    \centering
    \subfloat[On-time rate]{
    \includegraphics[width=0.45\linewidth]{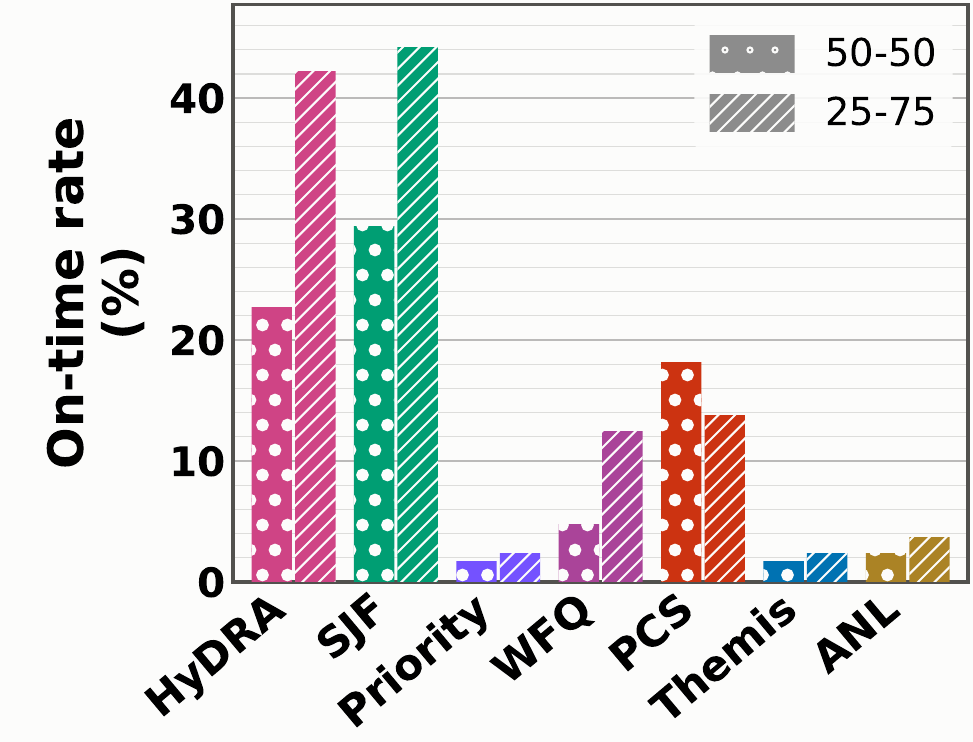}\label{fig:pair-large-ontime}}
    \subfloat[Project Jain's index]{
    \includegraphics[width=0.45\linewidth]{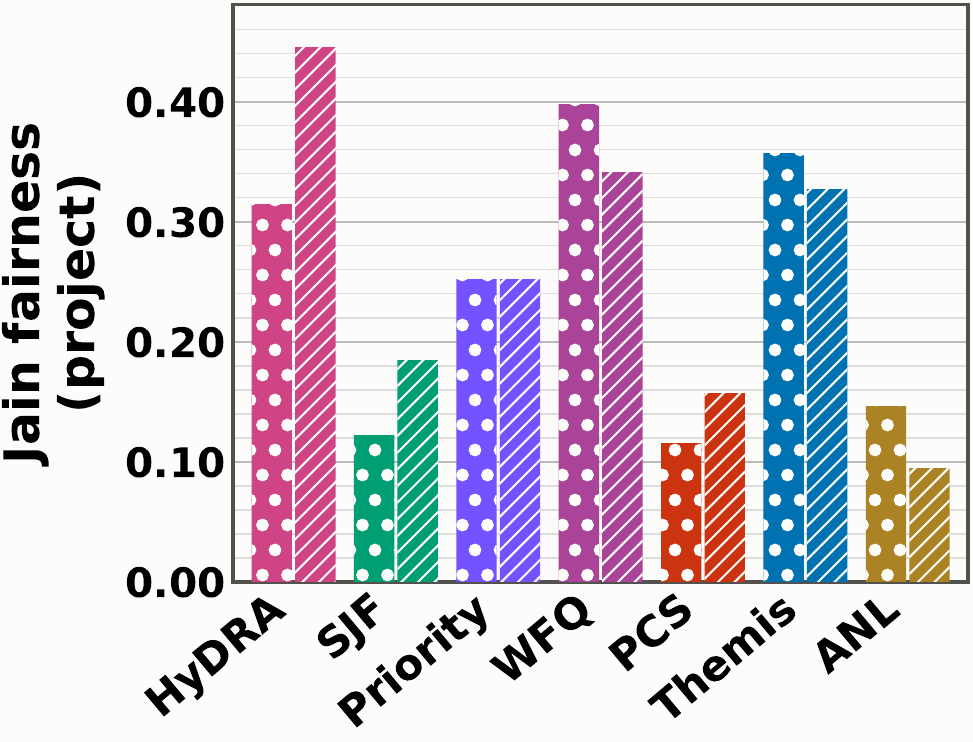}\label{fig:pair-large-jfi}}\\
    \subfloat[Project service lag (p95)]{
    \includegraphics[width=0.45\linewidth]{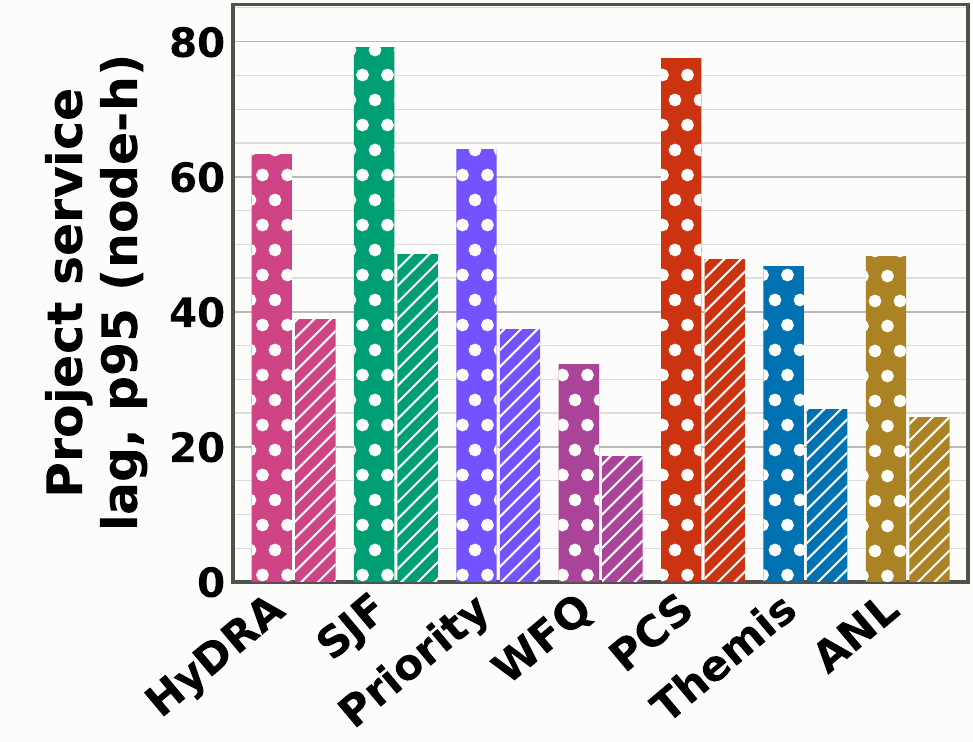}\label{fig:pair-large-lag}}
    \subfloat[Bounded slowdown (p95). Log scale.]{
    \includegraphics[width=0.45\linewidth]{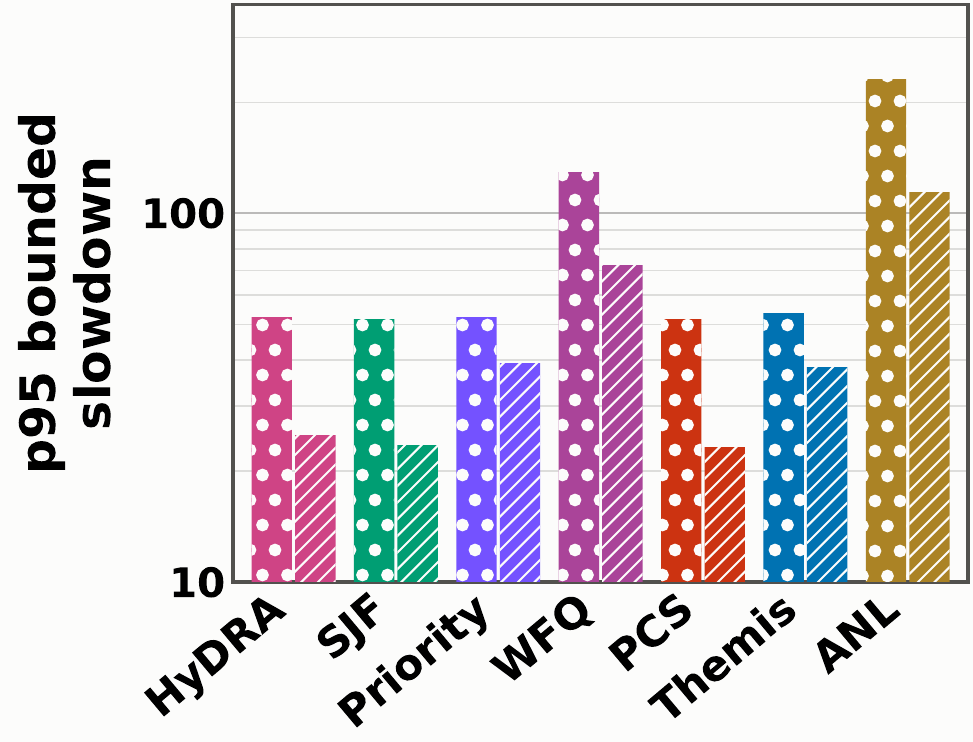}\label{fig:pair-large-bsd}}
    \caption{Comparison on varying demand conditions, 2 hour, Large workload, 25\% high, 75\% low using dots and 50-50 using hatch. }
    \label{fig:pair-large}
\end{figure}

Besides how many jobs arrive, contention also depends on how much each job demands, since a few high demand jobs can hold back many small ones. To study this effect without changing the activity or burstiness of the workload, we condition the demand of the 2\,h Large workload (l=0). Its job arrivals and priorities stay the same, but each job is made either high-demand, using 16 nodes for a walltime of 3,600\,s, or low-demand, using 1 node for 900\,s, and its runtime is set to 90\% of its walltime. We generate two such workloads, in which 50\% (\textit{50-50}) and 25\% (\textit{25-75}) of the jobs are high-demand, and compare the schedulers on both in Fig.~\ref{fig:pair-large}.

With more high-demand jobs, the 50-50 workload loads the cluster more heavily, so every scheduler waits longer: compared to 25-75, the project lag grows by 1.6--2.0$\times$ and the tail slowdown by up to 2.2$\times$ (Figs.~\ref{fig:pair-large-lag} and~\ref{fig:pair-large-bsd}). HyDRA is the fairest scheduler on 25-75, with a project-level Jain's index of 0.45 against 0.34 for WFQ and 0.33 for Themis, but WFQ (0.40) and Themis (0.36) overtake it (0.31) on 50-50 (Fig.~\ref{fig:pair-large-jfi}). Since runtimes here are a fixed 90\% of walltimes, the tags that WFQ computes from requested walltimes match the service that jobs actually receive, so WFQ comes close to the weighted fair share against which the index is measured. On the Polaris traces, where walltimes greatly overestimate runtimes, WFQ is far less fair (Fig.~\ref{fig:baseline-large-jfi}). WFQ also keeps the lowest project lag on both workloads.

HyDRA and SJF start the most jobs by their deadlines on both workloads, 42\% and 44\% on 25-75 and 23\% and 29\% on 50-50, whereas WFQ and Themis start at most 13\% (Fig.~\ref{fig:pair-large-ontime}). HyDRA's tail slowdown is also within 8\% of the best scheduler, while that of WFQ is 2.5--2.9$\times$ higher and that of ANL 4.4--4.6$\times$ higher. Overall, as high-demand jobs grow from 25\% to 50\% of the workload, HyDRA is the only scheduler that offers the best balance between fairness and deadlines. 

\section{Conclusion}
In this paper, we presented HyDRA, a hierarchical scheduler that shares an HPC cluster fairly among projects, users, and jobs while bounding the queue wait of every job. HyDRA assigns each job a deadline on arrival from its size, its priority, and the current backlog and ranks waiting jobs by a score that grows with their waiting time.
On workloads sampled from ALCF Polaris traces, HyDRA achieves $2\times$ fairness scores and lowers the 95th-\%ile bounded slowdown by 4.2$\times$ relative to the next best baseline, at a moderate cost in utilization.
As future work, we will adapt the deadlines to sustained overload and integrate HyDRA into a production resource manager such as Slurm.

\section*{Acknowledgments}
The authors would like to thank members of the Globus Labs at the University of Chicago, including Akshat Singhania, Augustus Ellerm and Haochen Pan for their assistance with this paper. They would also like to thank Parmanand Sinha, and the University of Chicago Research Computing Center (RCC), for their assistance and feedback. We also acknowledge the Argonne Leadership Computing Facility (ALCF) for access to the Polaris traces used in our evaluation.

\balance
\bibliographystyle{unsrt}
\bibliography{references}

\end{document}